\documentclass[pra,aps,twocolumn]{revtex4}
\usepackage{amsmath}
\usepackage{epsf}

\newcommand\fra[2]{{\textstyle{\frac{#1}{#2}}}}

\begin{document}

\title{Two-mode optical state truncation and generation of
maximally entangled states in pumped nonlinear couplers}

\author{Adam Miranowicz and Wies\l{}aw Leo\'nski}
 \affiliation{Nonlinear Optics Division, Institute of Physics, Adam
 Mickiewicz University, 61-614 Pozna\'n, Poland}

\date{\today}

\begin{abstract}
Schemes for optical-state truncation of two cavity modes are
analyzed. The systems, referred to as the nonlinear quantum
scissors devices, comprise two coupled nonlinear oscillators (Kerr
nonlinear coupler) with one or two of them pumped by external
classical fields. It is shown that the quantum evolution of the
pumped couplers can be closed in a two-qubit Hilbert space spanned
by vacuum and single-photon states only. Thus, the pumped couplers
can behave as a two-qubit system. Analysis of time evolution of
the quantum entanglement shows that Bell states can be generated.
A possible implementation of the couplers is suggested in a pumped
double-ring cavity with resonantly enhanced Kerr nonlinearities in
an electromagnetically-induced transparency scheme. The fragility
of the generated states and their entanglement due to the standard
dissipation and phase damping are discussed by numerically solving
two types of master equations.
\end{abstract}

\maketitle

\pagenumbering{arabic}

\section{Introduction}

Methods for preparation and manipulation of nonclassical states of
light have become an important research area in quantum optics
\cite{state}, especially in relation to possible optical
implementations of quantum computers and systems for quantum
communication and quantum cryptography \cite{Nie00}. Among various
schemes for optical-qubit generation, the so-called {\em quantum
scissors} device of Pegg {\em et al.} \cite{Peg98} produces a
superposition of vacuum and single-photon states,
$c_0|0\rangle+c_1|1\rangle$, by optical-state truncation of an
input single-mode coherent light. The Pegg {\em et al.} quantum
scissors device was studied in numerous papers (see, e.g.,
\cite{Vil99,Kon00,Par00,Ozd01,Villas01,Mir04,Mir05}), and tested
experimentally by Babichev {\em et al.} \cite{Bab02} and Resch
{\em et al.} \cite{Res02}. This simple scheme and its
generalizations for truncation of an input optical state to a
superposition of $d$ Fock states (the so-called qudits)
\cite{Villas01,Mir05} are based on linear optical elements, and
thus referred to as the {\em linear quantum scissors} devices.
Optical-state `truncation' can also be achieved in systems
comprising nonlinear elements (e.g., Kerr media)
\cite{Leo97,Dar00}, and thus will be referred to as the {\em
nonlinear quantum scissors} devices. The above-mentioned schemes
are restricted to the single-mode optical truncation. Here, by
generalizing our former scheme \cite{Leo04}, we present a
realization of nonlinear quantum scissors for optical-state
truncation of two cavity modes by means of a pumped nonlinear
coupler.

Two-mode nonlinear couplers have become, shortly after their
introduction by Jensen \cite{Jen82} and Maier \cite{Mai82}, one of
the important topics of photonics due to their wide potential
applications and relative simplicity (see, e.g., reviews
\cite{Sny91,Per00}). Among various types of the nonlinear optical
couplers, those based on Kerr effect have attracted especial
interest both in classical \cite{Jen82,Mai82,Sny91,Gry01} and
quantum \cite{Che96,Hor89,Kor96,Fiu99,Ibr00,Ari00,San03,ElOrany05}
regimes. The Kerr nonlinear couplers can exhibit variations of
self-trapping, self-modulation and self-switching effects. In
quantum regime, they can also generate sub-Poissonian and squeezed
light \cite{Hor89,Kor96,Fiu99,Ibr00,Ari00}. Possibilities of
entanglement generation were also studied in nonlinear couplers
operating by means of the Kerr effect \cite{San03,ElOrany05} or
degenerate parametric down-conversion \cite{Her03}.

Here, we analyze Kerr nonlinear couplers, which can be modelled by
systems composed of two quantum nonlinear oscillators linearly
coupled to each other and placed inside a double-ring cavity. We
discuss two schemes based on the coupler with an external
excitation of a single mode and the coupler with two modes pumped.
We show that the states generated in the excited nonlinear
couplers under suitable conditions can be limited to a
superposition of only vacuum and single-photon states,
$c_{00}|00\rangle+c_{01}|01\rangle+c_{10}|10\rangle+c_{11}|11\rangle$.
We compare the possibilities of generation of maximally entangled
states by the couplers excited in single and two modes. We also
discuss effects of dissipation on the fidelity of truncation and
suggest a method to achieve strong Kerr interactions at low
intensities in our system.

\begin{figure} 
\epsfxsize=6cm\centerline{\epsfbox{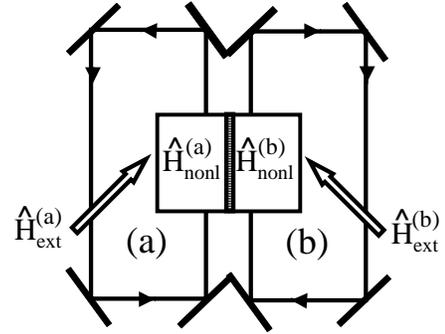}} \caption{A
realization of the two-mode nonlinear quantum scissors device via
a pumped nonlinear coupler in a double-ring cavity. Symbols are
explained in the text.}
\end{figure}

\section{Coupler pumped in a single mode}

We consider a system, referred to as the pumped Kerr nonlinear
coupler, which consists of two nonlinear oscillators, with Kerr
nonlinearities $\chi_a$ and $\chi_b$, linearly coupled to each
other and additionally linearly coupled to an external classical
field. In this section, we assume that the field is coupled to one
of the oscillators only. Thus, the system can be described by the
following Hamiltonian \cite{Leo04} ($\hbar=1$):
\begin{eqnarray}
\hat{H}&=&\hat{H}_{0}+\hat{H}_1,
\nonumber \\
\hat{H}_0&=&\omega_a\hat{a}^\dagger\hat{a}+\omega_b\hat{b}^\dagger\hat{b},
\nonumber \\
\hat{H}_1&=&\hat{H}^{(a)}_{\rm nonl} + \hat{H}^{(b)}_{\rm
nonl}+\hat{H}_{\rm int}+\hat{H}^{(a)}_{\rm ext}, \label{N01}
\end{eqnarray}
where
\begin{eqnarray}
\hat{H}^{(a)}_{\rm nonl}+\hat{H}^{(b)}_{\rm nonl} &=&
\frac{\chi_a}{2}(\hat{a}^\dagger )^2\hat{a}^2 +
\frac{\chi_b}{2}(\hat{b}^\dagger )^2\hat{b}^2, \label{N02}
\end{eqnarray}
\begin{eqnarray}
\hat{H}_{\rm int}&=&\epsilon\hat{a}^\dagger\hat{b} +\epsilon^*
\hat{a}\hat{b}^\dagger,
\label{N03} \\
\hat{H}^{(a)}_{\rm
ext}&=&\alpha\hat{a}^\dagger+\alpha^*\hat{a}, \label{N04}
\end{eqnarray}
and $\hat{a}$ ($\hat{b}$) is the bosonic annihilation operator
corresponding to the mode $a$ ($b$) of frequency $\omega_a$
($\omega_b$). Hamiltonian (\ref{N01}) for $\hat{H}^{(a)}_{\rm
ext}=0$ describes the standard nonlinear coupler
\cite{Che96,Hor89,Kor96,Fiu99,Ibr00,Ari00,San03,ElOrany05}
composed of two Kerr nonlinear oscillators linearly coupled to
each other, where the parameter $\epsilon$ is the strength of this
coupling. However, our model differs from the standard by
inclusion of the extra term $\hat{H}^{(a)}_{\rm ext}$, which
describes linear coupling between the driving single-mode
classical field (say, with frequency $\omega_{\rm ext}^{(a)}$) and
the cavity mode $a$. The parameter $\alpha$ describes the strength
of this coupling and is proportional to the classical field
amplitude.  A possible physical realization of the model is
presented in figure 1, where an external pump, described by
$\hat{H}^{(b)}_{\rm ext}$, is off in the present analysis. Kerr
media, marked by $\hat{H}^{(a)}_{\rm nonl}$ and
$\hat{H}^{(b)}_{\rm nonl}$, are linearly coupled as described by
grey central region corresponding to $\hat{H}_{\rm int}$.

The evolution of our system in the interaction picture can be
described by the Schr\"odinger equation
\begin{equation}
i\frac{d }{d t}|\psi (t)\rangle =\hat{H}_{1}|\psi (t)\rangle,
\label{N05}
\end{equation}
where
\begin{equation}
|\psi (t)\rangle =\sum_{m,n=0}^\infty c_{mn}(t)|mn\rangle .
\label{N06}
\end{equation}
Thus, the complex probability amplitudes $c_{mn}(t)$ satisfy the
set of equations of motion:
\begin{eqnarray}
&& i\frac{d }{d t}c_{mn}  = \left[ \chi _{a}m(m-1)+\chi
_{b}n(n-1)\right]c_{mn} \notag \\  && + \epsilon c_{m-1,n+1}
\sqrt{m(n+1)} + \epsilon ^{\ast } c_{m+1,n-1} \sqrt{(m+1)n} \notag
\\ &&  + \alpha c_{m-1,n} \sqrt{m} + \alpha ^{\ast } c_{m+1,n}
\sqrt{m+1}. \label{N07}
\end{eqnarray}
A superficial analysis of (\ref{N07}) could lead to a conclusion
that the evolution of the system pumped by classical external
field cannot be restricted to two lowest photon-numer states, but
will also include states with a greater number of photons.
However, by generalizing the method of single-mode nonlinear
quantum scissors proposed in \cite{Leo97} (for a review see
\cite{MLI01,LM01}), we have observed in \cite{Leo04} that
evolution can be restricted to the only four states: $|00\rangle$,
$|10\rangle$, $|01\rangle$ and $|11\rangle$ as a result of
degeneracy of Hamiltonians $\hat{H}^{(a)}_{\rm nonl}$ and
$\hat{H}^{(b)}_{\rm nonl}$. By assuming that the couplings
$|\alpha|$, and $|\epsilon|$ are much smaller than the Kerr
nonlinearities $\chi_a$ and $\chi_b$, we can interpret the
evolution between the four states as resonant transitions, while
the negligible evolution to other states as out of resonance,
analogously to the single-mode case \cite{LM01}. This phenomenon
can be shown explicitly as follows: under the assumption of $\chi
_{a},\chi _{b}\gg \max (|\epsilon |,|\alpha |)$ and short
evolution times, equation (\ref{N07}) for $m,n\neq 0,1$ can be
approximated by
\begin{equation}
i\frac{d }{d t}c_{mn}\approx \left[ \chi _{a}m(m-1)+\chi
_{b}n(n-1)\right] c_{mn} \label{N08}
\end{equation}
which has the simple solution
\begin{equation}
c_{mn}(t)\approx \exp \{-i\left[ \chi _{a}m(m-1)+\chi
_{b}n(n-1)\right] t\}c_{mn}(0). \label{N09}
\end{equation}
By setting the initial condition $c_{mn}(0)=0$ for $m,n\neq 0,1$,
one gets $c_{mn}(t)\approx 0$. By contrast, for $m,n\in \{0, 1\}$,
the terms proportional to $\chi_a$ and $\chi_b$ are vanishing due
to degeneracy of the Kerr Hamiltonian and so the remaining terms
proportional to $\epsilon$ and $\alpha$ are significant. Thus, the
ideally `truncated' two-mode state generated in the system has the
following simple form
\begin{equation}
|\psi (t)\rangle_{\rm cut} =c_{00}(t)| 0 0\rangle +c_{01}(t)| 0
1\rangle +c_{10}(t)| 1 0\rangle+c_{11}(t)| 1 1\rangle, \label{N10}
\end{equation}
where the evolution of $c_{mn}$, precisely given by (\ref{N07}),
can approximately be described by the following equations
\begin{eqnarray}
i\frac{d}{dt}c_{00}&=&\alpha^*c_{10},\quad
i{\frac{d}{dt}}c_{01}\;=\;\epsilon^*c_{10}+\alpha^*c_{11},\nonumber\\
i\frac{d}{dt}c_{11}&=&\alpha\, c_{01},\quad\;
i\frac{d}{dt}c_{10}\;=\;\epsilon\, c_{01}+\alpha\,
c_{00}.\label{N11}
\end{eqnarray}
Hereafter, in equations for the probability amplitudes under the
discussed assumptions, sign `=' should be understood as `$\approx
$'. Although approximate equations (\ref{N11}) are independent of
$\chi_a$, our derivation clearly shows that the Kerr nonlinearity
plays a crucial role in the truncation process. By assuming that
both oscillators are initially in vacuum states, $|\psi
(t=0)\rangle =| 0 0\rangle$, and parameters $\alpha$ and
$\epsilon$ are real, we find the following solutions of
(\ref{N11}) for the time-dependent probability amplitudes:
\begin{eqnarray}
c_{00}&=&\frac{1}{2\gamma}\left[(\gamma-\epsilon)\cos\tau
_{1}+(\gamma+\epsilon)\cos\tau _{2}\right],
\notag \\
c_{01}&=&\frac{\alpha}{\gamma}\left(\cos\tau _{1}-\cos\tau
_{2}\right),
\notag \\
c_{10}&=&-\frac{i(\gamma+\epsilon)\Omega_2}{4 \alpha
\gamma}(\sin\tau _{1}+\sin\tau _{2}),
\notag \\
c_{11}&=&-\frac{i}{2\gamma}\left(\Omega_2\sin\tau
_{1}-\Omega_1\sin\tau _{2}\right), \label{N12}
\end{eqnarray}
where
$\Omega_{j}=\sqrt{2[2\alpha^2+\epsilon^2+(-1)^{j-1}\epsilon\gamma]}$,
$\gamma=\sqrt{4\alpha ^{2}+\epsilon ^{2}}$, and $\tau _{j}=\Omega
_{j}t/2$ for $j=1,2$. Note that the solution for $c_{10}$ can be
written in a more symmetric form since the properties hold:
$(\gamma+\epsilon)\Omega_2=(\gamma-\epsilon)\Omega_1=2
\sqrt{\alpha^2(\gamma^2-\epsilon^2)}$. In a special case of equal
couplings $\alpha$ and $\epsilon$, (\ref{N12}) simplifies to our
former solution \cite{Leo04}:
\begin{eqnarray}
c_{00} &=&\cos (\sqrt{5}\tau)\cos (\tau)+\frac{1}{\sqrt{5}}\sin
(\sqrt{5}\tau)\sin (\tau), \notag
\\
c_{01} &=&-\frac{2}{\sqrt{5}}\sin (\sqrt{5}\tau)\sin (\tau), \notag \\
c_{10} &=&-i\frac{2}{\sqrt{5}}\sin (\sqrt{5}\tau)\cos (\tau), \label{N13} \\
c_{11} &=&-i\cos (\sqrt{5}\tau)\sin (\tau)+\frac{i}{\sqrt{5}}\sin
(\sqrt{5}\tau)\cos (\tau), \notag
\end{eqnarray}
where $\tau=\alpha t /2$. To estimate the quality of the
optical-state truncation (up to single-photon states) of the
generated light, we apply fidelity as a measure of discrepancy
between the ideally truncated two-qubit state $\hat{\rho}_{\rm
cut}=|\psi (t)\rangle_{\rm cut}\,_{\rm cut}\langle \psi (t)|$,
given by (\ref{N10}), and the actually generated output state
$\hat{\rho}(t)=|\psi(t)\rangle\langle\psi(t)|$ calculated
numerically from
\begin{eqnarray}
 |\psi(t)\rangle=\exp(-i\hat{H}t)|00\rangle
\label{N14}
\end{eqnarray}
for a large (practically infinite-dimensional) two-mode Hilbert
space. Specifically, in our numerical analysis, we have chosen the
dimension equal to $20$ for each subspace associated with single
mode of the field. The fidelity, also referred to as the Uhlmann's
transition probability for mixed states, is defined by (see, e.g.,
\cite{Ved02}):
\begin{figure} 
\epsfxsize=8cm\centerline{\epsfbox{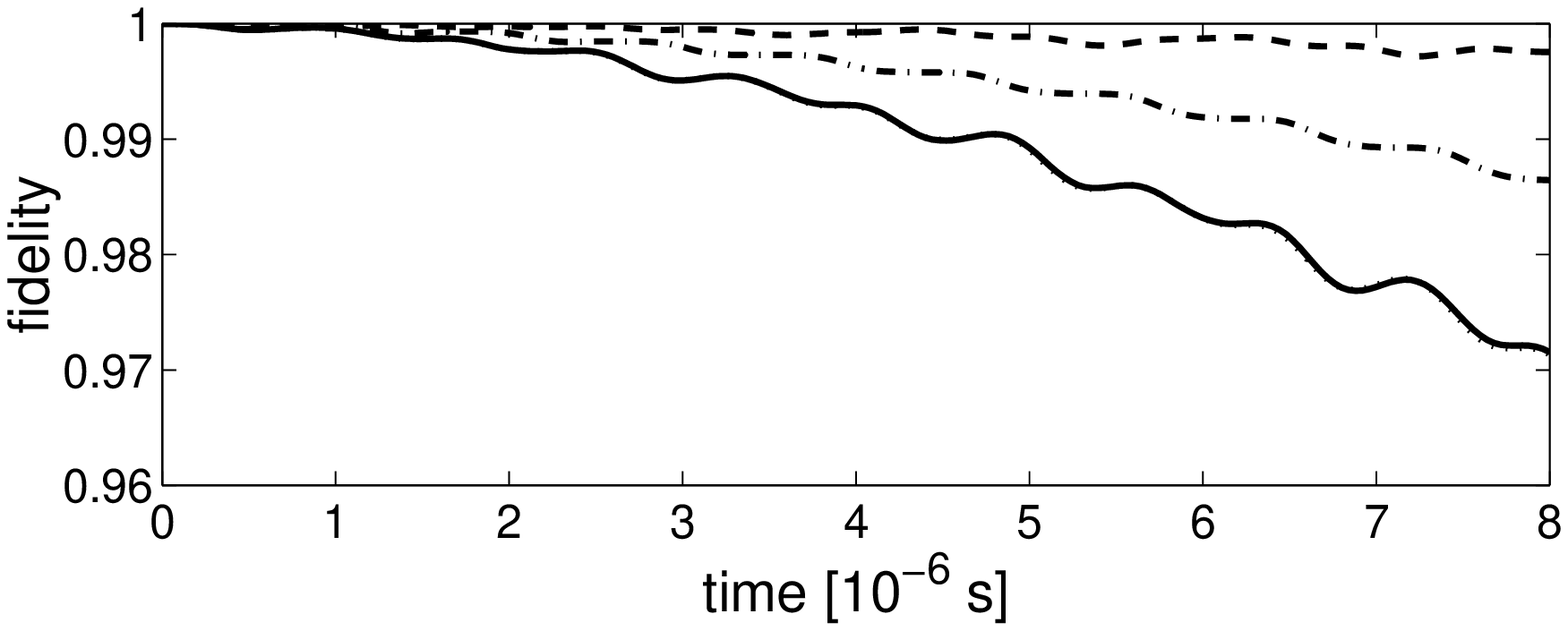}}
\epsfxsize=8cm\centerline{\epsfbox{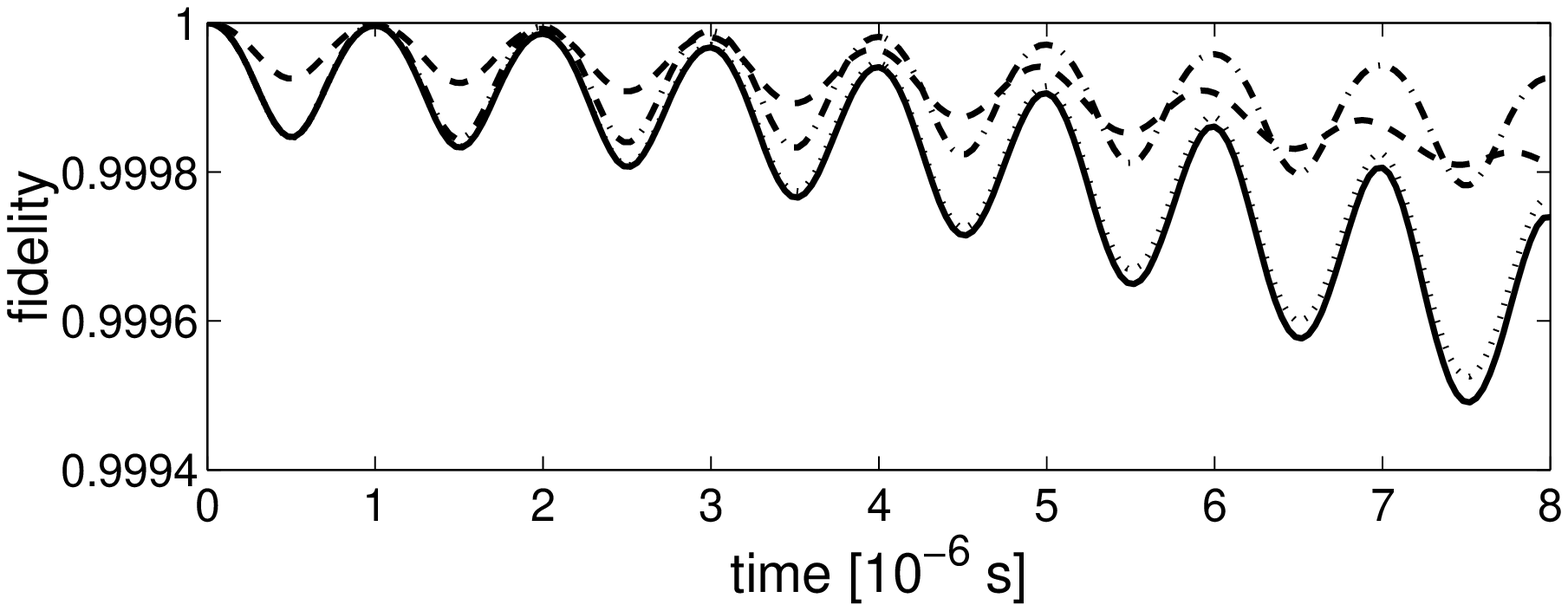}}%
\caption{Fidelity between the actually generated state
$|\psi(t)\rangle$ and the ideally truncated state $|\psi_{\rm cut
} (t)\rangle$ if the coupling strengths are (a) $\epsilon=\alpha$
and (b) $\epsilon=\alpha/10$ for the system pumped in one mode
with $\beta=0$ (dashed curves), and that pumped in two modes with:
$\beta=\alpha$ (solid), $\beta=-\alpha$ (dotted), $\beta=i\alpha$
(dot-dashed curves). In figures 2--7, we assume that the
nonlinearity coefficients are $\chi_a=\chi_b=10^8$ rad/sec,
$\alpha=\chi_a/200$, and the coupler is initially in two-mode
vacuum state $|00\rangle$.}
\end{figure}
\begin{figure} 
\centerline{ \epsfxsize=4.4cm\epsfbox{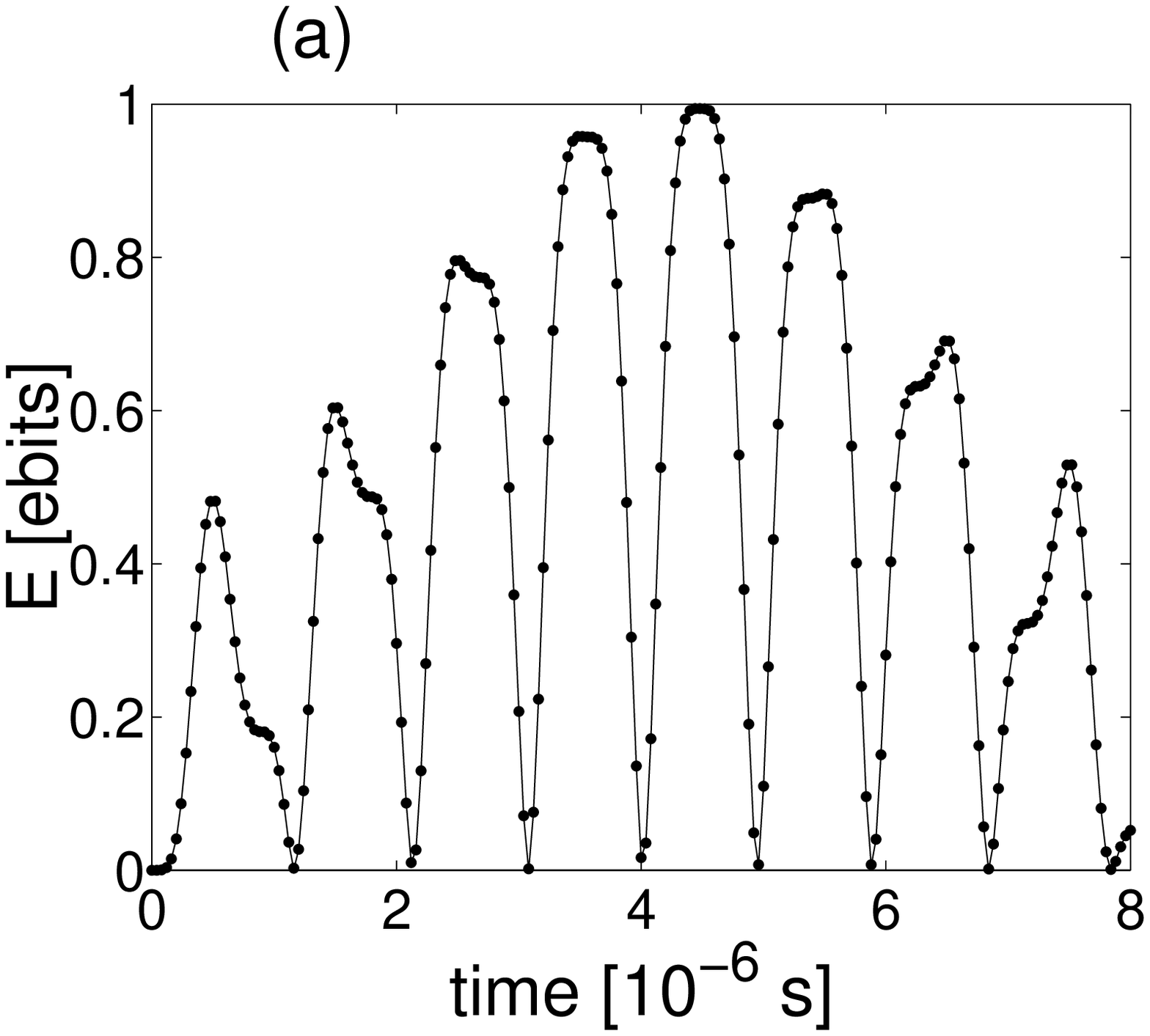}
\epsfxsize=4.4cm\epsfbox{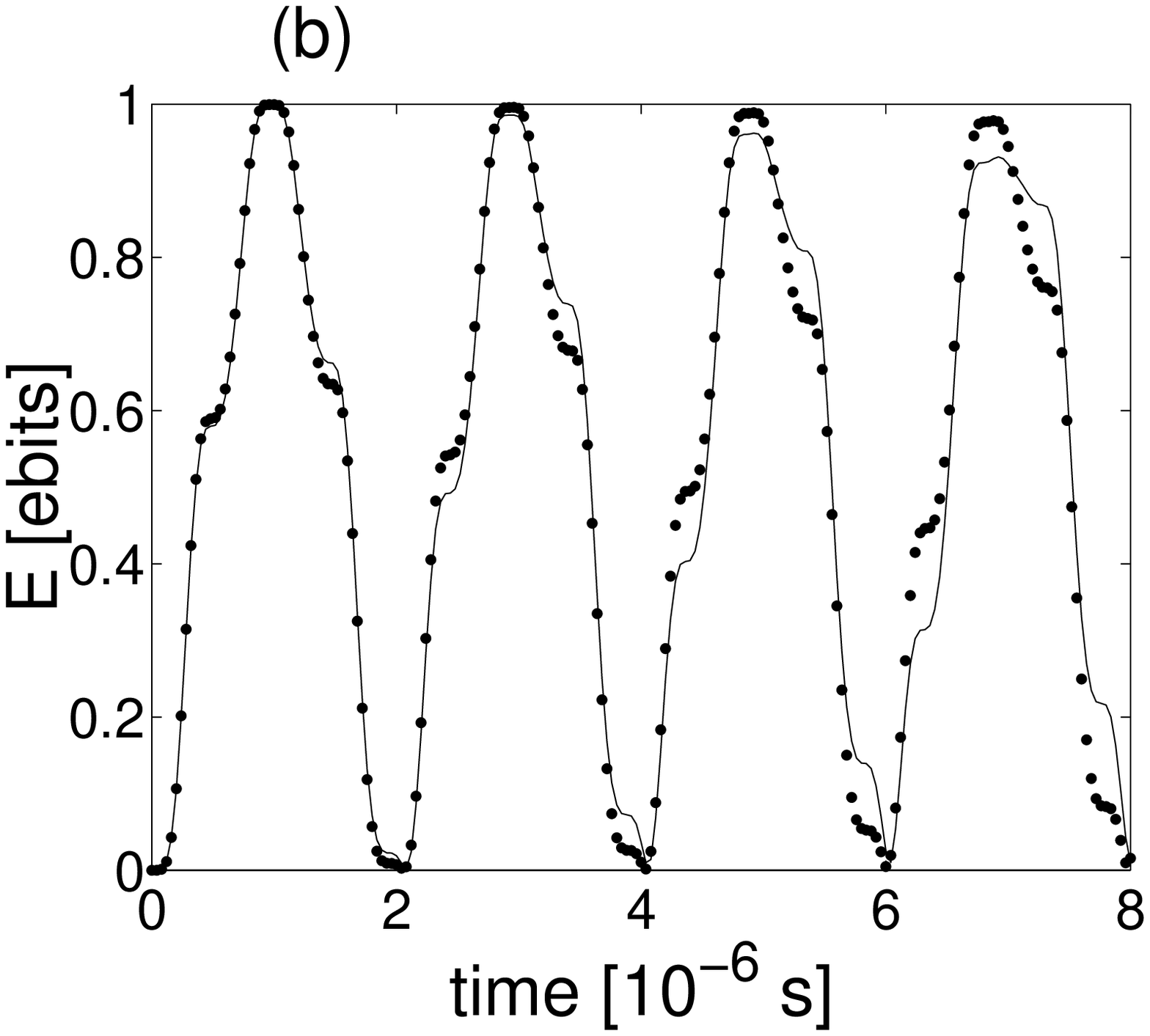}} %

\centerline{ \epsfxsize=4.4cm\epsfbox{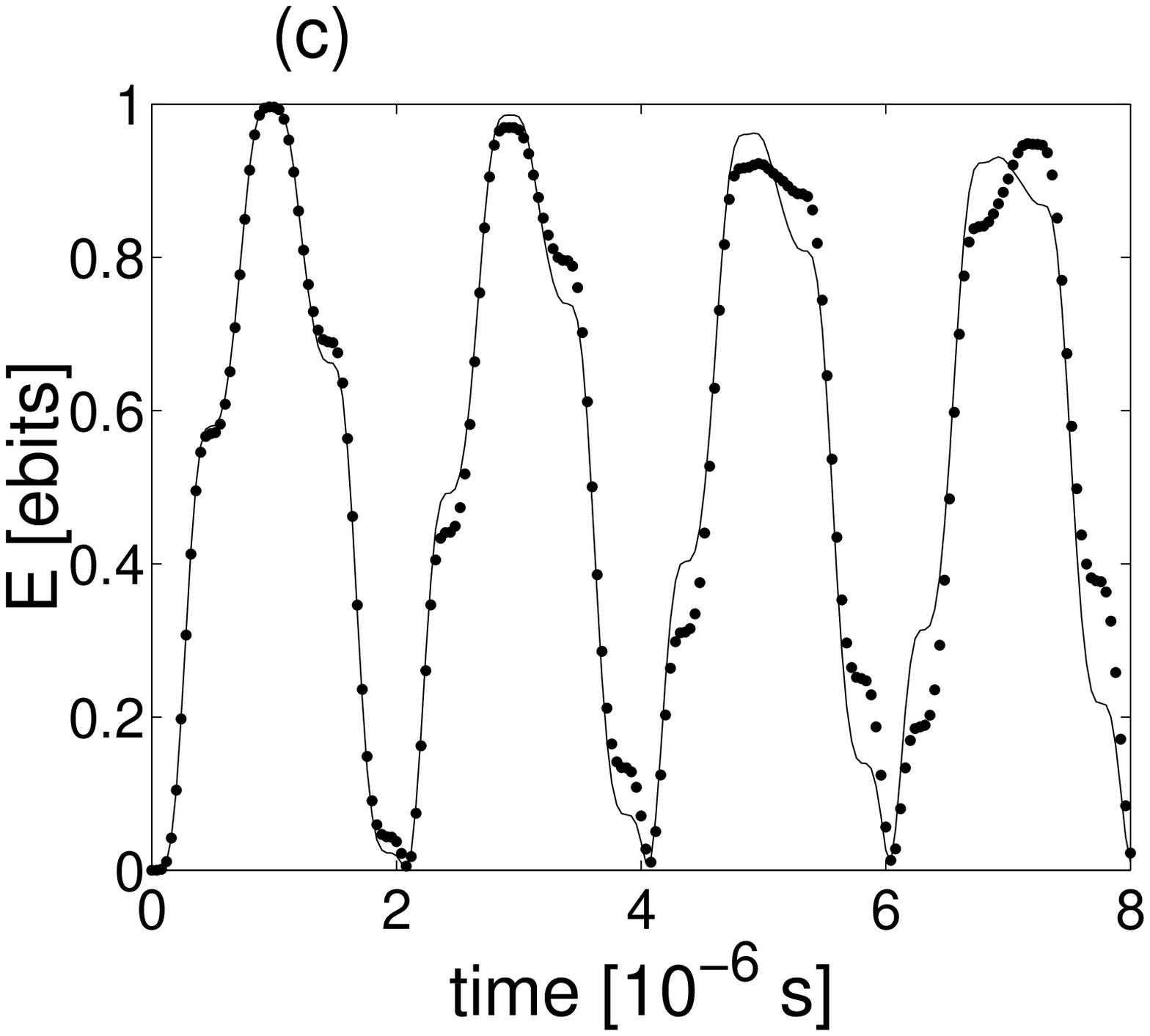}
\epsfxsize=4.4cm\epsfbox{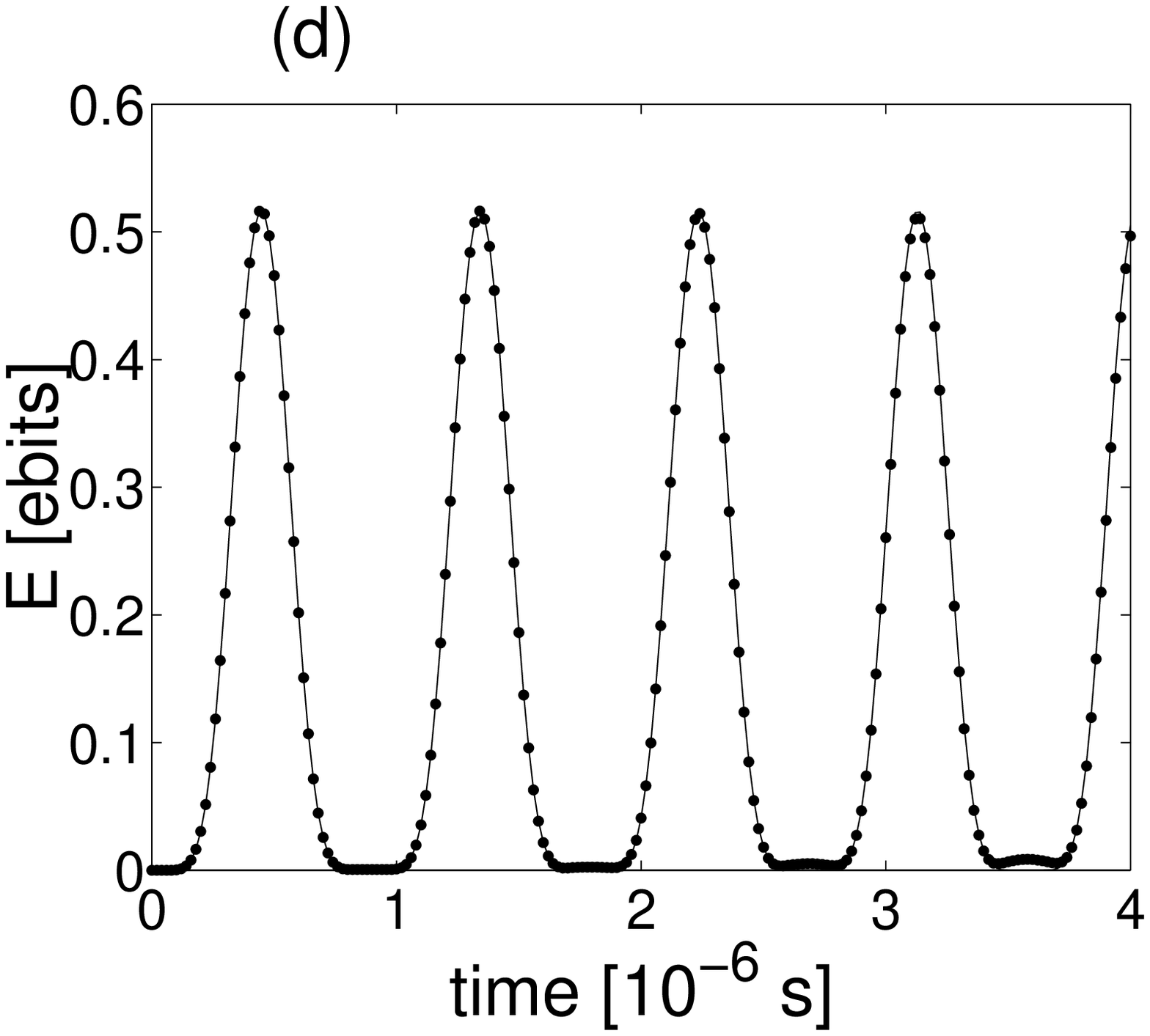}} %
\caption{Evolution of the entropy of entanglement $E$ of the
generated states $|\psi(t)\rangle$ (dots) and the desired
truncated states $|\psi_{\rm cut }(t)\rangle$ (solid curves) by
the coupler pumped in a single mode with (a) $\beta=0$ and in
two modes with: (b) $\beta=\alpha$, (c) $\beta=-\alpha$, (d)
$\beta=i\alpha$, where $\alpha=\epsilon$.}
\end{figure}
\begin{figure} 
\centerline{\epsfxsize=4.4cm\epsfbox{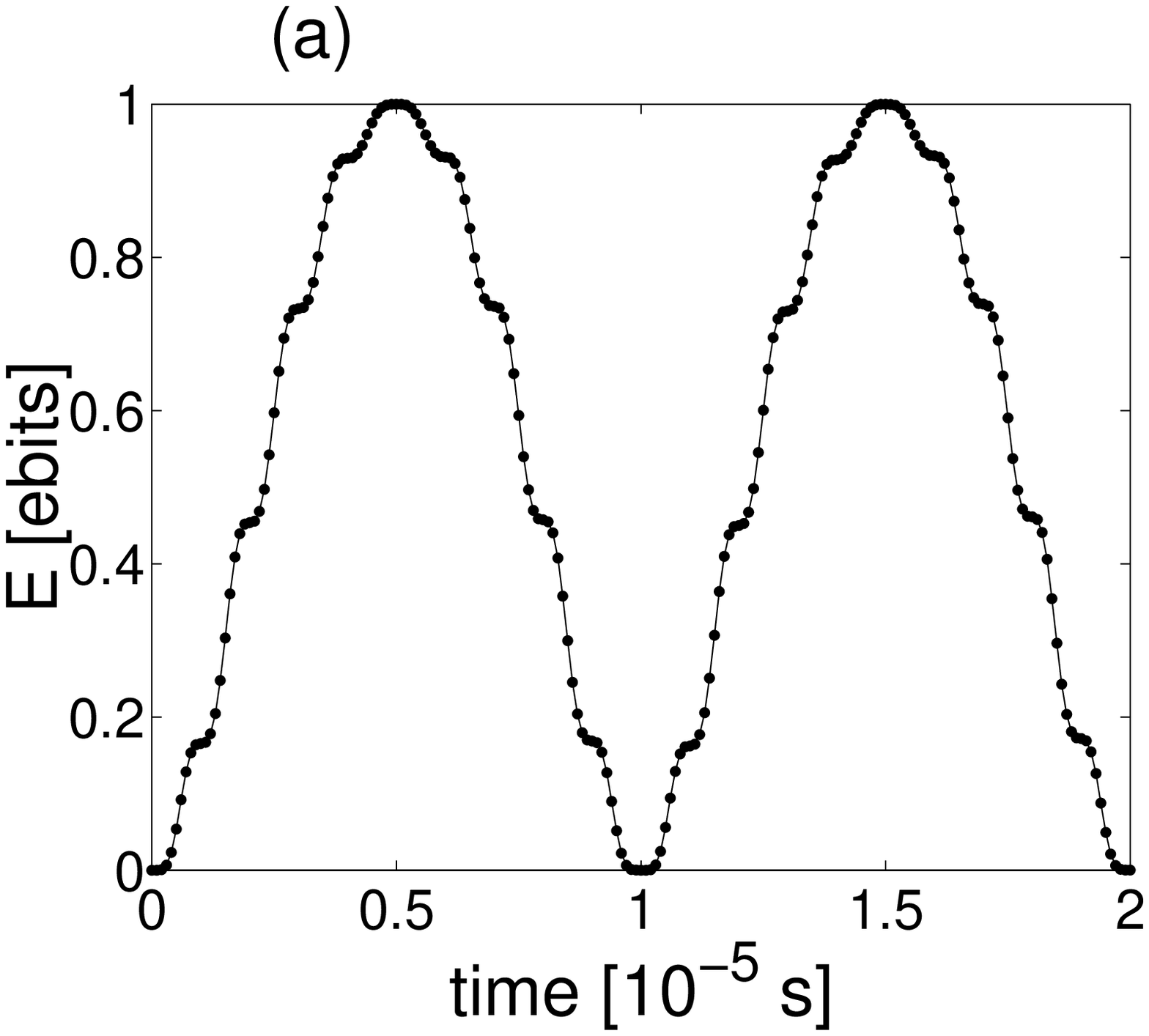}
\epsfxsize=4.4cm\epsfbox{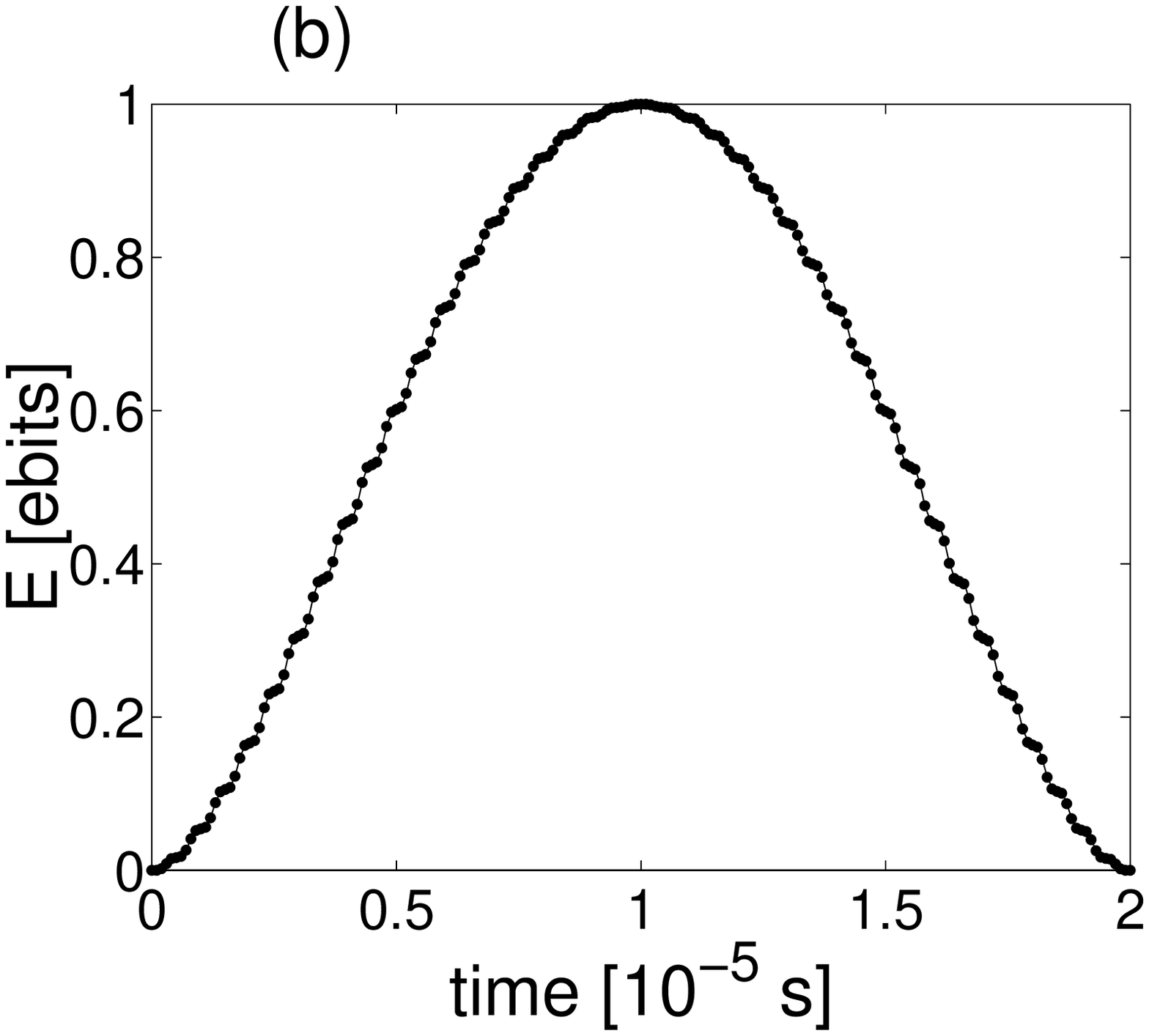}} %
\caption{Evolution of the entropy of entanglement $E$ for the
coupler pumped in (a) single mode ($\beta=0$) and (b) two
modes ($\beta=\alpha$) using line styles and parameters same as
in figure 3 except $\epsilon=\alpha/10$. On this figure scale,
evolutions of $E$ for $\beta=-\alpha$ can be described by the same
curve as in figure (b), while the entropy of entanglement for
$\beta=i\alpha$ is negligible as less than $5\times 10^{-4}$.}
\end{figure}
\begin{eqnarray}
F(\hat{\rho},\hat{\rho}_{\rm cut}) &=& \left\{ {\rm
Tr}[(\sqrt{\hat{\rho}_{\rm cut}}\hat{\rho} \sqrt{\hat{\rho}_{\rm
cut}})^{1/2}]\right\}^2.
 \label{N15}
\end{eqnarray}
By assuming that one of the states is pure (say $| \psi
\rangle_{\rm cut}$), then (\ref{N15}) simplifies to $F=\,_{\rm
cut}\langle \psi | \hat{\rho} | \psi \rangle_{\rm cut}$. Instead
fidelity, the Bures distance $D_B(\hat{\rho}|| \hat{\rho}_{\rm
cut}) = 2-2 \sqrt{F(\hat{\rho},\hat{\rho}_{\rm cut})}$ is often
applied as a measure of discrepancy between the states. Note that
the Bures distance satisfies the usual metric properties including
symmetry $D_B(\hat{\rho}||\hat{\rho}_{\rm
cut})=D_B(\hat{\rho}_{\rm cut}||\hat{\rho})$, contrary to the
quantum Kullback-Leibler `distance' (quantum relative entropy)
often used in quantum-information context \cite{Ved02}. General
expression (\ref{N15}) will be applied in our description of the
effect of damping on the optical-state truncation in section 4. In
this section, we are focused on pure-state truncation, for which
equation (\ref{N15}) simplifies to the familiar expression
$F=||\langle \psi (t)|\psi (t)\rangle_{\rm cut} ||^2$, where the
ideally truncated state $|\psi (t)\rangle_{\rm cut}$ is given by
(\ref{N10}) and the actually generated state $|\psi (t)\rangle$ is
calculated numerically from (\ref{N14}). The fidelity for perfect
truncation is equal to one. Figure 2 clearly indicates that the
fidelity of truncation using the pumped coupler is close to one
for short times and the coupling strengths much smaller than the
nonlinearity parameters ($|\alpha|,|\epsilon|\ll \chi_{a,b}$). The
numerical results shown in figure 2 confirm the validity of our
analytical approach at least for short evolution times. Thus, we
can refer to the system as a kind of nonlinear (as operating by
means of Kerr nonlinearity) quantum scissors device.

Solutions for probability amplitudes of the truncated states
enable a simple calculation of quantum entanglement, which is one
of the most fundamental resources of quantum information theory
\cite{Nie00}. It is well known that the entanglement of a
bipartite pure state, described by a density matrix $\hat \rho=
|\psi\rangle\langle\psi|$, can be described by the von Neumann
entropy of either the reduced density matrix $\hat \rho_a={\rm
Tr}_b \hat \rho$ or $\hat \rho_b={\rm Tr}_a \hat \rho$ or,
equivalently, by the Shannon entropy of the squared Schmidt
coefficients $p_i$ \cite{Ved02}:
\begin{eqnarray}
E(\hat \rho)&=&-{\rm Tr}(\rho_a\log_2\rho_a)=-{\rm
Tr}(\rho_b\log_2\rho_b)
\notag\\
&=&-\sum_{i=1}^N p _i\log_2 p_i \equiv h(p_1,...,p_{N-1}).
\label{N16}
\end{eqnarray}
For bipartite pure states this measure is often referred to as the
{\em entropy of entanglement}. In a special case of two qubits in
a pure state, the entropy of entanglement $E$ ranges from zero for
a separable state to 1 ebit for a maximally entangled state, and
it is simply given in terms of the binary entropy $h(p)= -p\log
_{2}p-(1-p)\log _{2}(1-p)$. In fact, for a general two-qubit pure
state, given by (\ref{N10}) with arbitrary amplitudes $c_{mn}$
($m,n=0,1$), the entropy of entanglement given by (\ref{N16}) can
simply be calculated as
\begin{equation}
 E(t) \equiv E(|\psi(t)\rangle_{\rm cut})
 = {\cal E}\left(2|c_{00}(t)c_{11}(t)-c_{01}(t)c_{10}(t)|\right),
\label{N17}
\end{equation}
where
\begin{eqnarray}
 {\cal E}(x) \equiv h\left(\frac{1}{2}(
1+\sqrt{1-x^{2}})\right) \label{N18}
\end{eqnarray}
and $h$ is the binary entropy. If the probability amplitudes
$c_{mn}(t)$ evolve according to (\ref{N12}) then the evolution of
the entropy of entanglement is given by
\begin{eqnarray}
E(t) = {\cal E}\Big( \sum_{j=1,2}\frac{\Omega_j}
{\gamma^{2}}\big\{ \epsilon\cos(\fra12 \Omega_{3-j}t)
-[\epsilon+(-1)^j\gamma]
\nonumber \\
 \times \cos(\fra12 \Omega_{j}t)\big \}
 \sin(\fra12 \Omega_{3-j}t)\Big). \label{N19}
\end{eqnarray}
Solution (\ref{N19}) is further simplified by assuming real
$\alpha=\epsilon$ then the amplitudes $c_{mn}$ are given by
(\ref{N13}). Thus, we obtain
\begin{eqnarray}
E (t)= {\cal E}\Big(\frac15 \{ [4+\cos (\sqrt{5}\alpha t)] \sin
(\alpha t)
\nonumber \\
 -\sqrt{5}\cos (\alpha t)\sin (\sqrt{5}\alpha t)\}\Big). \label{N20}
\end{eqnarray}
Figures 3(a) and 4(a) show the plots of the entropy of
entanglement for the case discussed here, i.e., for the
single-mode pumped coupler with the coupling parameters $|\alpha|$
and $|\epsilon|$ much smaller than the nonlinearities
$\chi_a=\chi_b$. We see in figure 3(a) that the rapid oscillations
in time (with a period $T_1=\pi/|\alpha|$) are modulated by
oscillations of low frequency (with period $T_2=8\pi/|\alpha|$).
As a consequence, their maxima are of various values but some of
them approach 1 ebit corresponding to the formation of Bell
states. To show this explicitly, we represent the generated state
in the basis $|\psi\rangle=\sum_{j=1}^4 b_j|B_j\rangle$ spanned by
the Bell-like states:
\begin{figure} 
\centerline{\epsfxsize=4.4cm\epsfbox{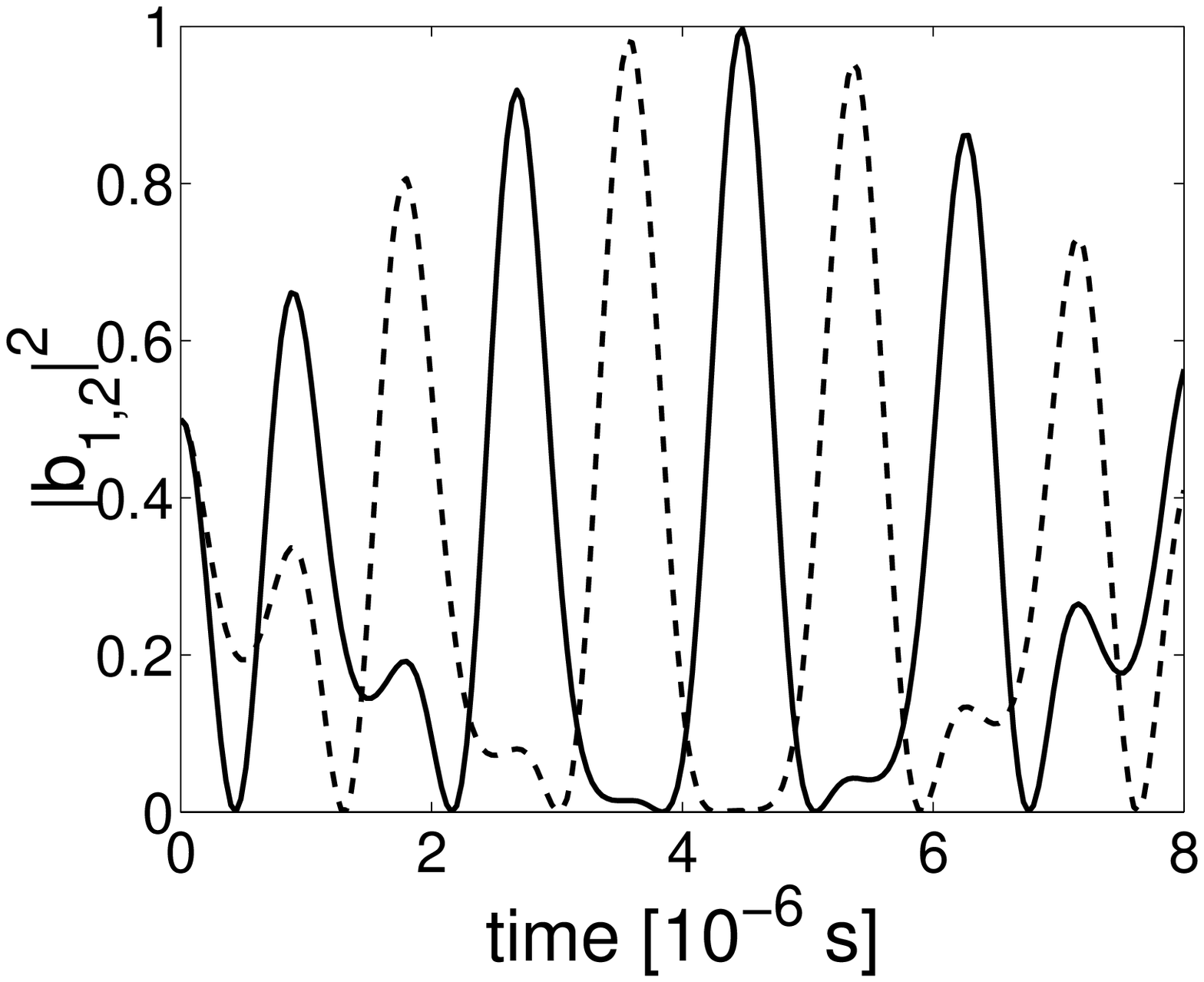}
\epsfxsize=4.4cm\epsfbox{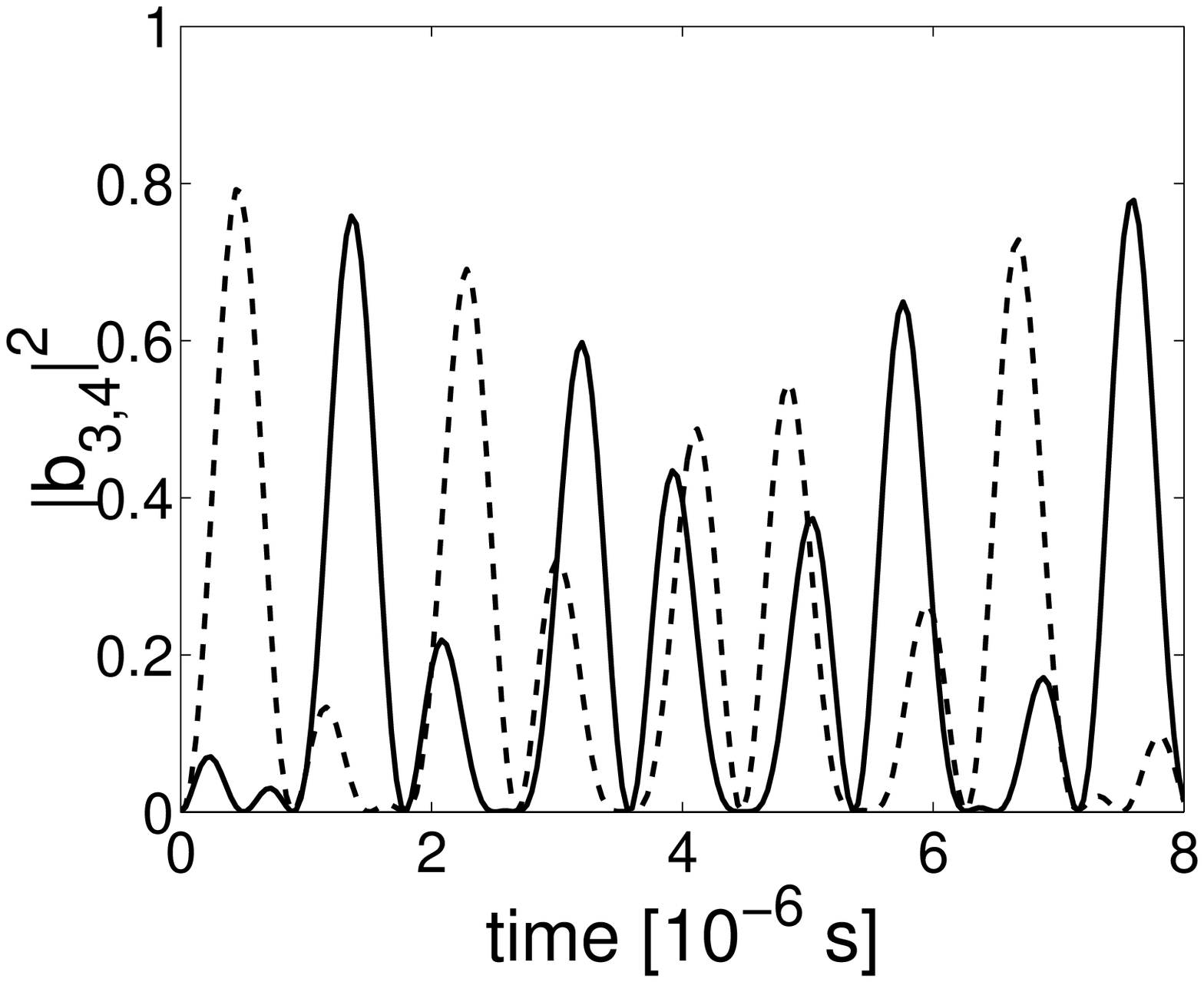}} %
\caption{Probabilities $|b_i|^2$ for finding the coupler pumped in
a single mode with $\epsilon=\alpha$ in the Bell states:
$|B_{1}\rangle$ and $|B_{3}\rangle$ (solid curves) as well as
$|B_{2}\rangle$ and $|B_{4}\rangle$ (dashed curves).}
\end{figure}
\begin{figure} 
\centerline{\epsfxsize=4.4cm\epsfbox{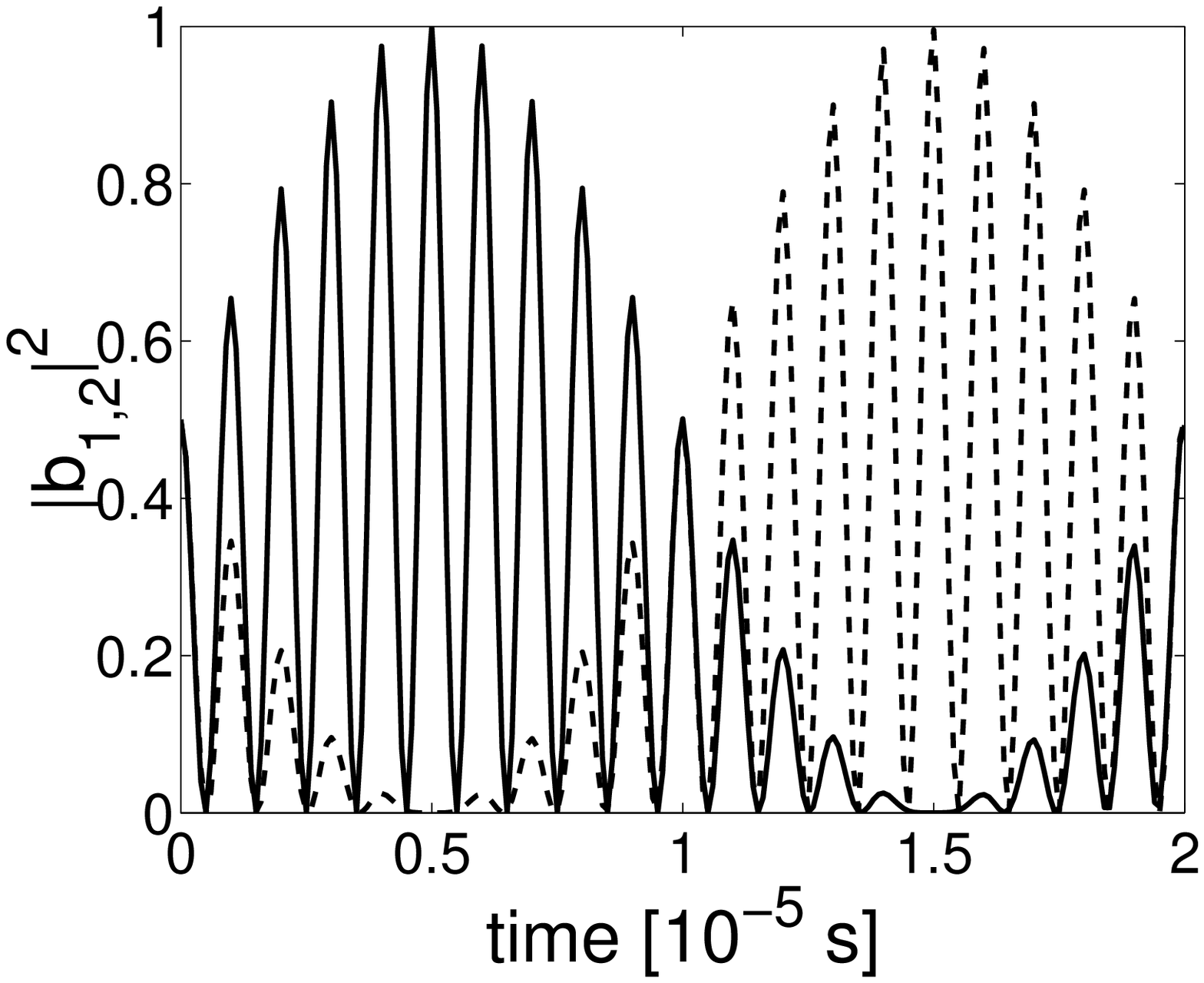}
\epsfxsize=4.4cm\epsfbox{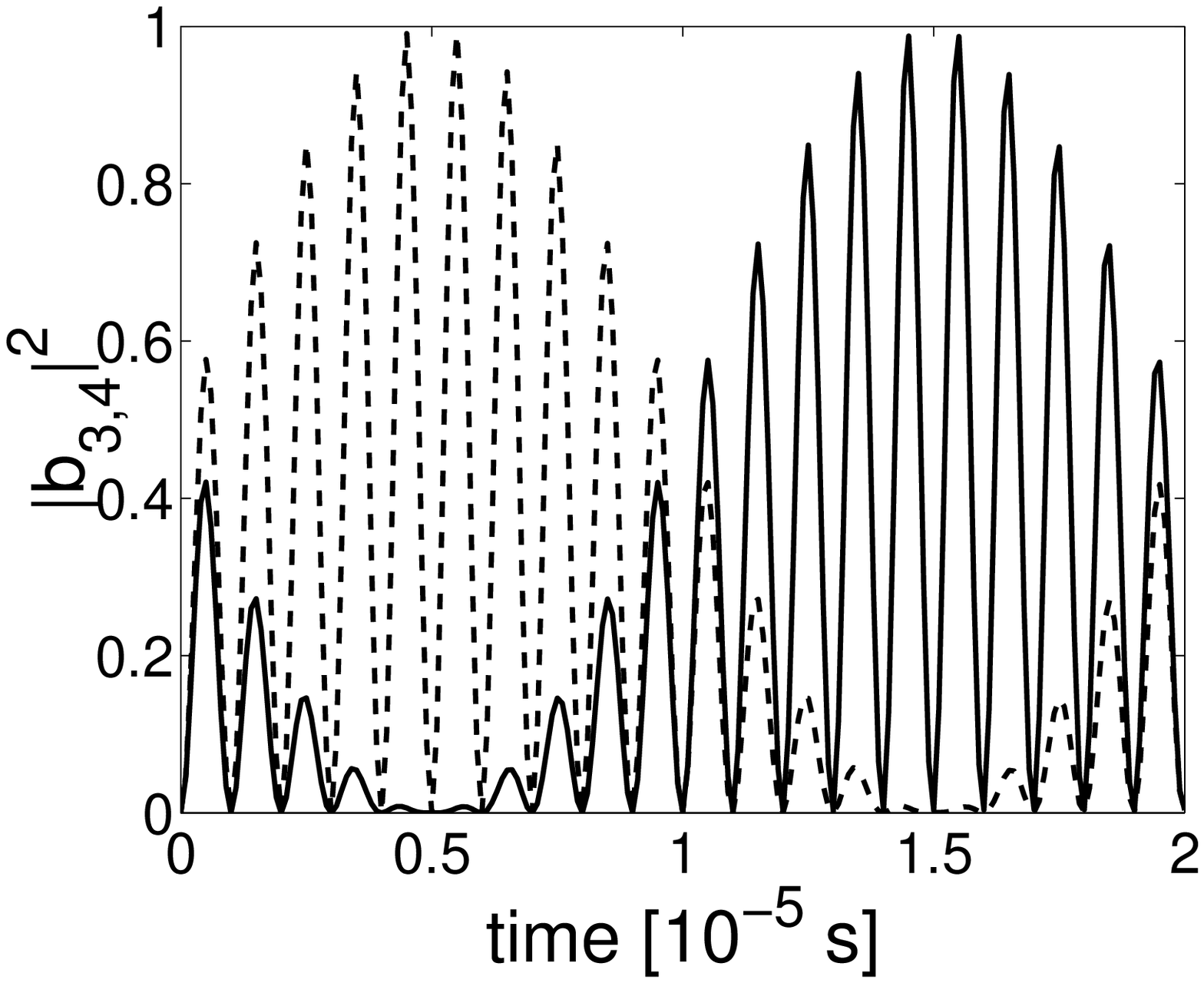}} %
\caption{Probabilities for finding the single-mode pumped
coupler with $\epsilon=\alpha/10$ in the Bell states. Same symbols
as in figure 5.}
\end{figure}
\begin{figure} 
\centerline{\epsfxsize=4.4cm\epsfbox{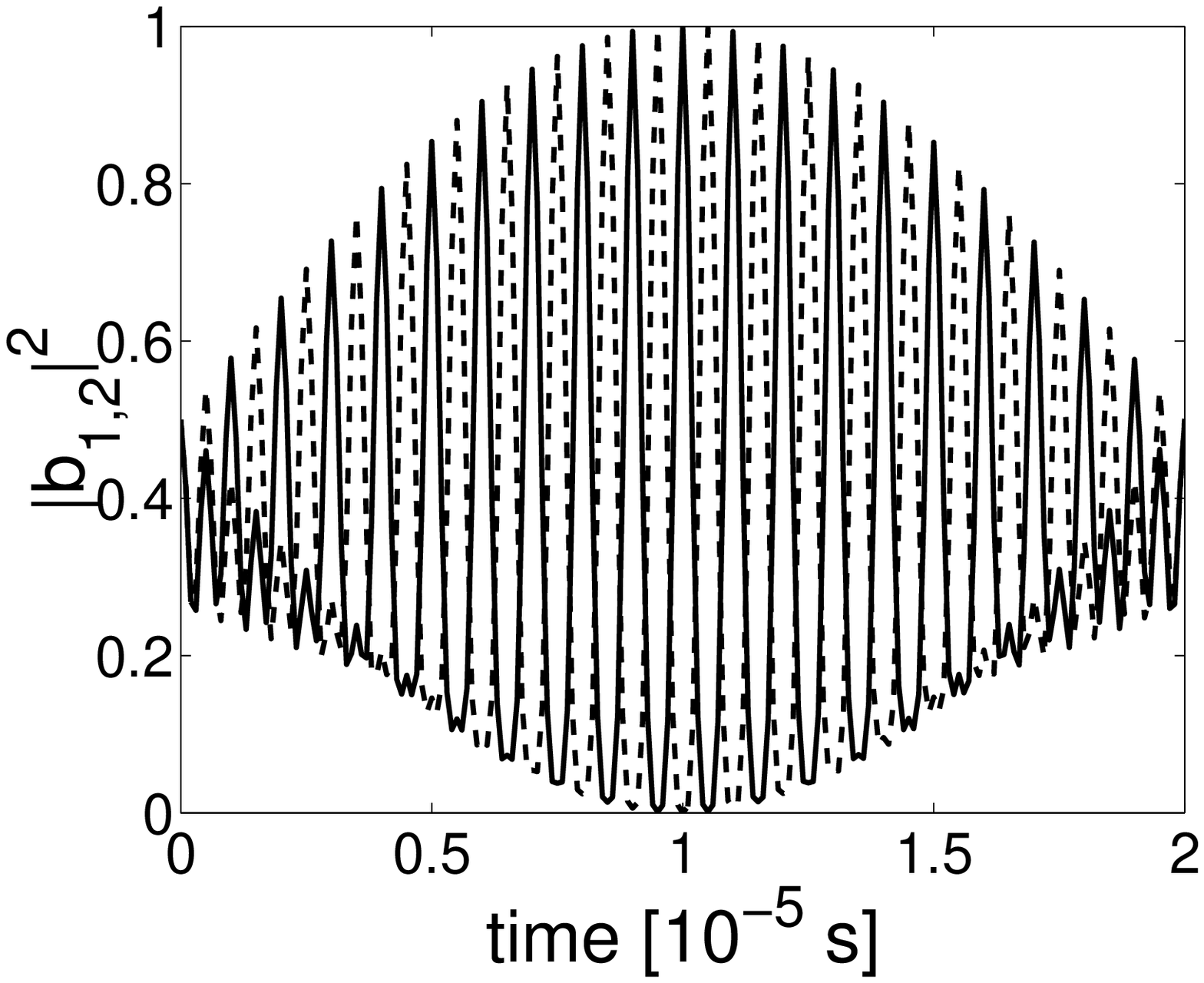}
\epsfxsize=4.4cm\epsfbox{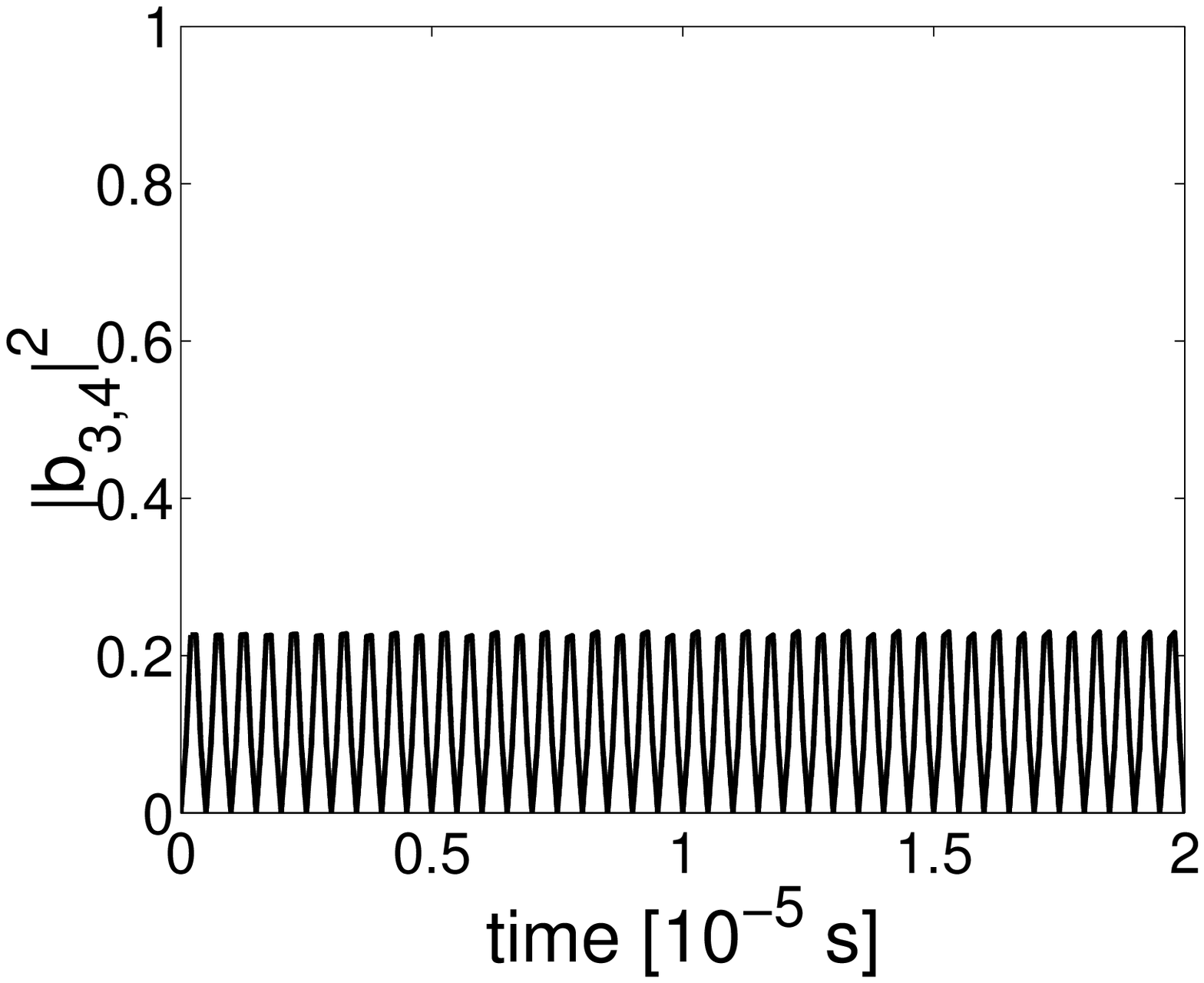}} %
\caption{Probabilities for finding the coupler pumped in two
modes with $\alpha=\beta$ and $\epsilon=\alpha/10$ in the Bell
states. Same symbols as in figure 5.}
\end{figure}
\begin{eqnarray}
|B_1\rangle&=&\frac{|11\rangle+i|00\rangle}{\sqrt{2}},\quad
|B_2\rangle\,=\,\frac{|00\rangle+i|11\rangle}{\sqrt{2}},\nonumber\\
|B_3\rangle&=&\frac{|01\rangle-i|10\rangle}{\sqrt{2}},\quad
|B_4\rangle\,=\,\frac{|10\rangle-i|01\rangle}{\sqrt{2}},
\label{N21}
\end{eqnarray}
which differ from the standard Bell states only by the phase
factor $i$. Clearly, $\sum_j |B_j\rangle\langle B_j|=1$ and
$E(|B_j\rangle)=1$ for $i=1,...,4$, so states (\ref{N21}) will
shortly be referred to as the Bell states. Figures 5 and 6 show
the probabilities for the generation of the Bell states as a
function of time for the single-mode driven couplers, when the
initial state is the two-mode vacuum state. It is seen that states
$|B_1\rangle$ and $|B_2\rangle$, being superpositions of
$|00\rangle$ and $|11\rangle$, are generated with a high accuracy.
In detail, the maxima of (\ref{N20}) occur approximately at times
$t(m,n) = \frac12 [(2m-1)T_1+(2n-1)T_2]$ if $\alpha=\epsilon$.
Since the frequencies of oscillations in (\ref{N20}) are
incommensurate, by waiting long enough (assuming no dissipation),
we can achieve 1 ebit with high precision. For example, in the
first four periods of $T_2$, one could observe generation of
entangled states approaching Bell state $|B_1\rangle$ with
$E[t(1,1)]=0.994$ and $E[t(1,2)]=0.997$ ebits, and approaching
$|B_2\rangle$ with $E[t(2,3)]=0.992$ and $E[t(2,4)]=0.998$. Note
that only the first period is shown in figures 3(a) and 5(a), for
which the highest entanglement of 0.994 ebits occurs approximately
at the time $t(1,1)=4.5 \times 10^{-6}$. Thus, we can effectively
treat our system as a source of maximally entangled states. On the
other hand, the Bell states $|B_3\rangle$ and $|B_4\rangle$, being
superpositions of $|10\rangle$ and $|01\rangle$, are not generated
if the initial state is vacuum and $\epsilon=\alpha$. This
conclusion can be drawn by observing that the probabilities for
$|B_3\rangle$ and $|B_4\rangle$ can reach only $0.8$ in figure
5(b). However, by relaxing the condition of equal couplings
$\epsilon$ and $\alpha$, as shown in figure 6(b), Bell states
$|B_3\rangle$ and $|B_4\rangle$ can be generated from vacuum with
high precision in the dissipation-free system.

Hitherto, we have assumed that both cavity modes were initially in
vacuum states. Now, we analyze a more general evolution when the
cavity modes are initially not only in vacuum but also in
single-photon Fock states, i.e., $|\psi (0)\rangle \equiv|\psi
^{(kl)}(0)\rangle =|kl\rangle$, where $k,l=0,1$. Thus, by assuming
as usual that $|\alpha|= |\epsilon|\ll\chi_a,\chi_b$, the
evolutions of the initial states are found to be
\begin{eqnarray}
|\psi ^{(01)}(\tau )\rangle_{\rm cut} &=&c_{01}|00\rangle
+\widetilde{c}_{00}|01\rangle +
\widetilde{c}_{11}|10\rangle +c_{10}|11\rangle , \notag \\
|\psi ^{(10)}(\tau )\rangle_{\rm cut} &=&c_{10}|00\rangle
+\widetilde{c}_{11}|01\rangle +
\widetilde{c}_{00}|10\rangle +c_{01}|11\rangle , \notag \\
|\psi ^{(11)}(\tau )\rangle_{\rm cut} &=&c_{11}|00\rangle
+c_{10}|01\rangle +c_{01}|10\rangle +c_{00}|11\rangle \notag \\
\label{N22}
\end{eqnarray}
in terms of the time-dependent amplitudes $c_{mn}(\tau )$ given by
(\ref{N13}) and
\begin{eqnarray}
\widetilde{c}_{00}(\tau ) =&\cos (\sqrt{5}\tau )\cos (\tau
)-\frac{1}{\sqrt{5}}
\sin (\sqrt{5}\tau )\sin (\tau ), \notag \\
\widetilde{c}_{11}(\tau ) =&-i\cos (\sqrt{5}\tau )\sin (\tau
)-\frac{i}{\sqrt{5}} \sin (\sqrt{5}\tau )\cos (\tau ), \label{N23}
\end{eqnarray}
where, as usual, $\tau=\alpha t/2$. Please note that
$\widetilde{c}_{jj}(\tau)$ and $c_{jj}(\tau)$ differ in sign. We
find that the generalized expression for the entropies of
entanglement for the initial states $|kl\rangle$ with $k,l=0,1$
reads as
\begin{eqnarray}
E^{(kl)}(t) = {\cal E}\Big(\frac15[4 +\cos (\sqrt{5}\alpha t)]
\sin (\alpha t)\hspace{1.5cm}
\nonumber \\
 -(-1)^{k-l}\sqrt{5}\cos (\alpha t)\sin (\sqrt{5}\alpha t)\Big),
\label{N24}
\end{eqnarray}
which implies that
\begin{eqnarray}
E^{(00)}(t)&=& E^{(11)}(t),
\notag \\
E^{(01)}(t)&=& E^{(10)}(t). \label{N25}
\end{eqnarray}
It is worth noting that our system for the initial states
$|01\rangle$ or $|10\rangle$ quasi-periodically evolves into the
Bell states $|B_3\rangle$ and $|B_4\rangle$ with high precision
but does not evolve into $|B_1\rangle$ or $|B_2\rangle$ assuming
$\alpha=\epsilon$. This is by contrast to the evolutions of the
initial states $|00\rangle$ (see figure 5) or $|11\rangle$ also
for $\epsilon=\alpha$. For brevity, we will not present any graphs
of the evolutions of the initial states $|01\rangle$,
$|10\rangle$, and $|11\rangle$, which would correspond to figures
2--5 plotted for the initial vacua.

The above solutions for the initial single-photon states are
included for completeness of our mathematical approach to show
that, in principle, all Bell states can be generated in our system
even for $\epsilon=\alpha$. But it should be stressed that the
system with the initial Fock states is much more experimentally
challenging than that assuming initially the vacuum states only.
Despite of experimental difficulty, our system enables generation
of the one-photon Fock states from vacuum assuming no coupling
$\epsilon$ between the modes, which corresponds to having two
independent pumped cavities with nonlinear Kerr media. A
possibility of producing single-photon states in such systems was
demonstrated in \cite{Leo97} (for a review see \cite{LM01} and
references therein).

\section{Coupler pumped in two modes}

This section is devoted to the most general scheme presented in
figure 1, namely that involving two external excitations.  We
assume here that both modes of the coupler are excited by external
fields, whereas for the case discussed previously we assumed that
only one of the modes was coupled to the external field. The
Hamiltonian describing such a system is of the form
\begin{equation}
\hat{H}_1=\hat{H}^{(a)}_{\rm nonl}+\hat{H}^{(b)}_{\rm
nonl}+\hat{H}_{\rm int} +\hat{H}^{(a)}_{\rm
ext}+\hat{H}^{(b)}_{\rm ext}, \label{N26}
\end{equation}
which is the same as that defined by (\ref{N01})--(\ref{N04}),
except for the extra term given by
\begin{eqnarray}
\hat{H}^{(b)}_{\rm ext}&=&\beta\hat{b}^\dagger+\beta^*\hat{b}
\label{N27}
\end{eqnarray}
corresponding to the coupling of the cavity mode $b$ with an
external driving single-mode classical field (say, with frequency
$\omega_{\rm ext}^{(b)}$), where the parameter $\beta$ describes
the strength of this interaction being proportional to the
classical field amplitude. The evolution of our system, described
by Hamiltonian (\ref{N26}), can be given by the Schr\"odinger
equation from which we find the following set of equations for the
amplitudes $c_{mn}(t)$ of the wavefunction (\ref{N06}) in the
interaction picture:
\begin{eqnarray}
&&  i\frac{d }{d t}c_{mn}  = \left[ \chi _{a}m(m-1)+\chi
_{b}n(n-1)\right]c_{mn}  \notag \\ && + \epsilon c_{m-1,n+1}
\sqrt{m(n+1)} +
\epsilon ^{\ast } c_{m+1,n-1} \sqrt{(m+1)n} \notag \\
&& + \alpha c_{m-1,n} \sqrt{m} + \alpha ^{\ast } c_{m+1,n}
\sqrt{m+1}\notag \\  && +\beta c_{m,n-1} \sqrt{n} +\beta ^{\ast }
c_{m,n+1} \sqrt{n+1}. \label{N28}
\end{eqnarray}
Analogously to the analysis in the former section, we assume short
evolution times as well as the couplings $|\alpha|$, $|\beta|$,
and $|\epsilon|$ to be much smaller than the Kerr nonlinearities
$\chi_a$ and $\chi_b$. Then, equation (\ref{N28}) for $m,n\neq
0,1$ can be approximated by (\ref{N08}) with the solution
(\ref{N09}), which vanishes for the initial condition
$c_{mn}(0)=0$. Thus, under the above assumptions, the infinite set
of equations (\ref{N28}) reduces to the following four equations:
\begin{eqnarray}
i\frac{dc_{00}}{dt}&=&\alpha^* c_{10}+\beta^*c_{01},\nonumber\\
i\frac{dc_{01}}{dt}&=&\epsilon^* c_{10}+\alpha^* c_{11}+\beta c_{00},\nonumber\\
i\frac{dc_{10}}{dt}&=&\epsilon c_{01}+\alpha c_{00}+\beta^*c_{11},\nonumber\\
i\frac{dc_{11}}{dt}&=&\alpha c_{01}+\beta c_{10}. \label{N29}
\end{eqnarray}
Assuming that at the time $t=0$, both oscillator modes are in the
vacuum states, i.e., $c_{00}=1$ and $c_{01}=c_{10}=c_{11}=0$, we
can obtain analytical solutions of (\ref{N29}). To solve
(\ref{N29}) we need to find zeros of a fourth-order polynomial
and, hence, the solutions in their general form are rather
complicated and unreadable. However, if we assume that all
coupling constants are real and the couplings with two external
fields are of the same strength ($\alpha=\beta$) then the
solutions become much simpler and easier to interpret. Thus, under
these assumptions, the solutions are found to be:
\begin{eqnarray}
c_{00} &=&\frac{1}{2}\left\{1+\left[\cos (\frac{\lambda
t}{2})+i\frac{\epsilon}{\lambda} \sin (\frac{\lambda
t}{2})\right]e^{-i\epsilon t/2}\right\}, \notag \\
c_{01} &=&c_{10}=-i2\frac{\alpha }{\lambda }\sin (\frac{\lambda
t}{2})\,e^{-i\epsilon t/2}, \notag \\
c_{11} &=&c_{00}-1, \label{N30}
\end{eqnarray}
where we have introduced the effective coupling constant
$\lambda=\sqrt{16\alpha^2+\epsilon^2}$. As in the case discussed
in the previous section, the system's dynamics is closed within
the finite set of $n$-photon states. Figure 2 shows that for the
assumed parameters and evolution times shorter than $5\times
10^{-6}$s, the fidelity between the ideally truncated states and
the actually generated states by means of the coupler pumped in
two modes deviates from 1 by the values less than $0.03$ in figure
2(a) or even $<6 \times 10^{-4}$ in figure 2(b). This again
confirms the validity of our analysis and justifies referring to
this system as a kind of the quantum scissors device. For the
parameters assumed in figure 2, truncation with higher fidelity is
usually observed for the coupler pumped in a single mode rather
than in two modes. In the latter case, the truncation fidelity
depends on the relative phase between the pumping-field couplings
$\alpha$ and $\beta$. Note that state $|\psi_{\rm cut}\rangle$ for
$\beta=\pm\alpha$, contrary to $\beta=i\alpha$, can be directly
calculated from solution (\ref{N12}). For brevity, we have not
presented here an analogous solution for $\beta=i\alpha$, but we
have used it for plotting the corresponding curves in figures 2
and 3(d). It is seen, by comparing figures 2(a) and 2(b), that by
decreasing $\epsilon$ in comparison to $\alpha$, the fidelity of
truncation can be improved for $\beta\neq 0$. Thus, pumping the
coupler in two modes can lead sometimes to truncation better than
that for the system driven in single mode as presented in figure
2(b) by dot-dashed curve corresponding to $\beta=i\alpha$ and
$\epsilon=\alpha/10$.

The evolution of the pure-state entanglement, given by
(\ref{N16}), generated in the coupler pumped in two modes can be
calculated from (\ref{N17}) with the probability amplitudes given
by (\ref{N30}). Thus, we obtain
\begin{equation}
E(t) = {\cal E}\Big(\frac1{2}\big|1-\frac{e^{-i\epsilon
t}}{\lambda^{2}}\big[16 \alpha^2 +\epsilon^2 \cos(\lambda t) +i
\epsilon \lambda \sin(\lambda t)\big]\big|\Big).\label{N31}
\end{equation}
Figures 3(b-d) and 4(b) show the evolution of the entropy of
entanglement measured in ebits as a function of time for the
system pumped in two modes in comparison to the results for the
single-mode driven coupler shown in figures 3(a) and 4(a). It is
seen that the first maximum in figures 3(b-d) is the highest,
contrary to the case shown in figures 3(a) for the single-mode
pumping. Nevertheless, the most important fact is that the value
of the entropy of entanglement $E$ can approach unity to a high
precision. So, as in the case of the single-mode pumping, the
two-mode driven system effectively generates Bell states. To find
which Bell states are generated, we can also transform the
resulting wavefunction into the Bell basis. Thus, figure 7 depicts
probabilities for the four Bell states. As expected from the form
of our analytical solutions for $c_{mn}$ ($m,n=0,1$), the
entanglement occurs for the states $|00\rangle$ and $|11\rangle$,
and leads to the generation of the states $|B_1\rangle$ and
$|B_2\rangle$ (with some unimportant global phase factor).
Clearly, the highest peaks of the entropy $E(t)$ and those of the
probabilities $|b_{1,2}(t)|^2$ of the Bell state generation occur
at the same evolution times, as seen by comparing figures: 3(a)
with 5(a), 4(a) with 6(a), and 4(b) with 7. Analysis of the
evolutions of the probabilities $|b_{1,2}(t)|^2$ and the entropy
of entanglement $E(t)$ for relatively longer times reveals that
the oscillations are modulated and some long-time oscillations
occur in the system, which can be interpreted as a result of
quantum beats. It is seen from (\ref{N31}) that two various
frequencies appear in our solution and one of them is considerably
greater than the other, e.g., the effective coupling constant
$\lambda$ is $\sqrt{17}$ times greater than the internal coupling
constant of the coupler $\epsilon=\alpha$. We are not presenting
here dissipation-free evolutions exhibiting modulated oscillations
at times longer than those in figure 4. It would be meaningless
since the entanglement is lost at such evolution times due to
dissipation, which inevitably occurs in real physical
implementations of the coupler as will be discussed in the next
section.

Although our analysis, including all figures, is focused on
evolution of the initial vacuum states, we present shortly some
results for other states too. We find that the evolution of the
initial Fock state $|\psi ^{(kl)}(0)\rangle =|kl\rangle$ with
$k,l=0,1$ is of the form (\ref{N22}) but with the probability
amplitudes $c_{mn}(\tau )$ given by (\ref{N30}) and
$\widetilde{c}_{mn}(\tau )$ equal to
\begin{eqnarray}
\widetilde{c}_{00}(\tau ) &=& \frac12 \Big\{e^{i\epsilon
t}+e^{-i\epsilon t/2} \Big[ \cos (\frac{\lambda t}{2})
-i\frac{\epsilon}{\lambda}
\sin (\frac{\lambda t}{2}) \Big] \Big\}, \notag \\
\widetilde{c}_{11}(\tau ) &=& \widetilde{c}_{00}(\tau
)-e^{i\epsilon t}. \label{N32}
\end{eqnarray}
With the help of these formulae, we can calculate the entropies of
entanglement explicitly as
\begin{eqnarray}
E^{(kl)}(t) ={\cal E}\Big( \frac1{2}
\Big|1-\lambda^{-2}\exp[-i(2|k-l|+1)\epsilon t] \hspace{0.5cm}
\nonumber \\
+\big[16 \alpha^2 +\epsilon^2 \cos(\lambda t)(-1)^{k-l}i \epsilon
\lambda \sin(\lambda t)\big]\Big|\Big) \label{N33}
\end{eqnarray}
implying the same properties as those given by (\ref{N25}) for the
single-mode excited system. We point out that both single- and
two-mode pumped couplers for $\alpha=\epsilon$ and the initial
states $|01\rangle$ or $|10\rangle$ evolve into the Bell states
$|B_3\rangle$ and $|B_4\rangle$ but do not evolve into
$|B_1\rangle$ or $|B_2\rangle$. This is contrary to the evolutions
of the initial vacuum states (or $|11\rangle$) as, for example,
presented in figures 5 and 7.

\section{Dissipation}

In a more realistic description both fields $a$ and $b$ lose their
photons from the cavities. According to the standard techniques in
theoretical quantum optics, dissipation of our system can be
modelled by its coupling to reservoirs (heat baths) as described
by the interaction Hamiltonian
\begin{eqnarray}
\hat H &=&  \hat H_1 + \hat{H}_{\rm loss}, \label{N34}
\\
\hat{H}_{\rm loss} &=& \hat \Gamma_a \hat f(\hat a,\hat{a}^{\dag
})+ \hat \Gamma_b \hat f(\hat b,\hat{b}^{\dag })+{\rm h.c.},
\label{N35}
\end{eqnarray}
where
\begin{eqnarray}
 \hat \Gamma_{a} = \sum_{j=0}^\infty g_{j}^{(a)}
\hat{c}_{j}^{(a)},\quad \hat \Gamma_{b} = \sum_{j=0}^\infty
g_{j}^{(b)} \hat{c}_{j}^{(b)}  \label{N36}
\end{eqnarray}
are the reservoir operators; $\hat{c}_{j}^{(a,b)}$ are the boson
annihilation operators of the reservoir oscillators coupled with
mode $a$ or $b$, respectively; $g_{j}^{(a,b)}$ are the coupling
constants of the interaction with the reservoirs; $\hat H_1$ is
given either by (\ref{N01}) or (\ref{N26}) dependent on the system
analyzed. We assume two kinds of functions $\hat f(\hat
a,\hat{a}^{\dag })$ and $\hat f(\hat b,\hat{b}^{\dag })$ to
describe standard damping and dephasing.

The standard description of a damped system is obtained for
(\ref{N35}) with $\hat f(\hat a,\hat{a}^{\dag })=\hat a^\dag$ and
$\hat f(\hat b,\hat{b}^{\dag })=\hat b^\dag$, or explicitly
\begin{eqnarray}
\hat{H}_{\rm loss} &=& \hat \Gamma_a\hat{a}^{\dag }+ \hat
\Gamma_a^{\dag }\hat{a}+ \hat \Gamma_b \hat{b}^{\dag }+\hat
\Gamma_b^{\dag } \hat{b}, \label{N37}
\end{eqnarray}
corresponding to energy transfer between the system and
reservoirs. It should be stressed that the process described by
(\ref{N37}) leads to combined effect of amplitude and phase
damping. The evolution in the Markov approximation of the reduced
density operator $\hat{\rho}$ of the two cavity modes after
tracing out over the reservoirs can be described in the
interaction picture by the following master equation (see, e.g.,
\cite{Gardiner})
\begin{eqnarray}
\frac{d \hat{\rho}}{d t }&=& -i[\hat{H}_1,\hat{\rho}]+ \hat {\cal
L}_{\rm loss}\hat \rho, \label{N38}
\end{eqnarray}
where the Liouvillian
\begin{eqnarray}
\hat {\cal L}_{\rm loss}\hat \rho = \frac{\gamma_a}{2}
 ([\hat a\hat \rho,\hat a^\dagger ]+[\hat a,\hat \rho\hat a^\dagger ])
 +\frac{\gamma_b}{2}
 ([\hat b\hat \rho,\hat b^\dagger ]
 \nonumber \\
 \quad +[\hat b,\hat \rho\hat b^\dagger ])
 +\gamma_a \bar n_a [[\hat a,\hat \rho],\hat a^\dagger]
 +\gamma_b \bar n_b [[\hat b,\hat \rho],\hat b^\dagger]
\label{N39}
\end{eqnarray}
is the usual loss term corresponding to $\hat{H}_{\rm loss}$,
given by (\ref{N37}); $\gamma_k$ is the damping rate of the $k$th
($k=a,b$) (ring) cavity, and $\bar n_k=[\exp(\hbar\omega_k/k_B
T)-1]^{-1}$ is the mean number of thermal photons at the reservoir
temperature $T$. We will analyze both `noisy'reservoirs (at $T>0$
implying $\bar n_a,\bar n_b> 0$) and `quiet' reservoirs (at
$T\approx 0$, so $\bar n_a=\bar n_b\approx 0$). The latter
assumption implies that diffusion of fluctuations from the
reservoirs into the system modes is negligible. But still this
simplified master equation describes the loss of photons from the
system modes to the reservoirs.

One can raise some doubts \cite{Alicki} against using Liouvillian
(\ref{N39}) in a description of lossy anharmonic oscillator models
given by (\ref{N02}). Also the approximation of $T=0$ in
(\ref{N39}) is problematic. Nevertheless, master equation
(\ref{N38}) with Liouvillian (\ref{N39}) and Hamiltonian $\hat
H_1$ set to $\hat{H}^{(a)}_{\rm nonl}$ was used in a number of
works both for $T>0$ (see, e.g., \cite{Dan89,Per90,Chat91,Tanas})
but also for $T=0$ (see, e.g.,
\cite{Mil86,Per88,Mil89,Dan89,Gardiner,Tanas}). Moreover, the same
master equation, given by (\ref{N38}) for $\hat H_1$ set to
(\ref{N02}), for coupled anharmonic oscillators was applied in,
e.g., \cite{Chat91,Per94}. Liouvillian (\ref{N39}) for $T=0$ was
also used in (\cite{Gardiner} p. 210) to describe a model
essentially similar to ours comprising a nonlinear system, in
which two quantized field modes in a cavity interact with a
classical pump field. The standard Liouvillian for $T=0$ was also
used in \cite{Mil91} to describe a system of Kerr nonlinearity,
given by $\hat{H}^{(a)}_{\rm nonl}$, and a parametric amplifier
driven by a pulsed classical field. Nevertheless, it should be
noted that a realistic master equation for Kerr medium
\cite{Dun99,Haus93} is more complicated.

Phase damping (also referred to as dephasing) can be described by
(\ref{N35}) assuming $\hat f(\hat a,\hat{a}^{\dag })=\hat{a}^{\dag
} \hat a$ and $\hat f(\hat b,\hat{b}^{\dag })=\hat{b}^{\dag }\hat
b$, which gives the following loss Hamiltonian \cite{Gardiner}:
\begin{eqnarray}
\hat{H}_{\rm loss} &=& ( \hat \Gamma_a
+\hat \Gamma_a^\dag )\hat{a}^{\dag } \hat{a} + ( \hat \Gamma_b
+\hat \Gamma_b^\dag )\hat{b}^{\dag } \hat{b} . \label{N40}
\end{eqnarray}
This interaction can be interpreted as a scattering process, where
the number of photons remains unchanged contrary to the
interaction described by (\ref{N37}). Phase damping is essential
in a fully quantum picture of dissipation of our system. As a
simple generalization of the Gardiner-Zoller master equation for a
single harmonic oscillator (\cite{Gardiner}, equation (6.1.15)),
we describe phase damping of our two-mode nonlinear system by
master equation (\ref{N38}) for the Liouvillian
\begin{eqnarray}
\hat {\cal L}_{\rm loss}\hat \rho = \frac{\gamma_a}{2} (2 \bar
n_a+1) [2\hat a^\dag \hat a \hat \rho \hat a^\dag \hat a - (\hat
a^\dag
\hat a)^2 \hat \rho-\hat \rho (\hat a^\dag \hat a)^2] \notag \\
\quad  + \frac{\gamma_b}{2} (2 \bar n_b+1) [2\hat b^\dag \hat b
\hat \rho \hat b^\dag \hat b - (\hat b^\dag \hat b)^2 \hat
\rho-\hat \rho (\hat b^\dag \hat b)^2], \label{N41}
\end{eqnarray}
which is significantly different from (\ref{N39}).

To analyze dissipative evolution, governed by master equations
(\ref{N38}) for Liouvillians (\ref{N39}) and (\ref{N41}), we
apply standard numerical procedures for solving ordinary
differential equations with constant coefficients as an
exponential series.
\begin{figure} 
\epsfxsize=7cm\centerline{\epsfbox{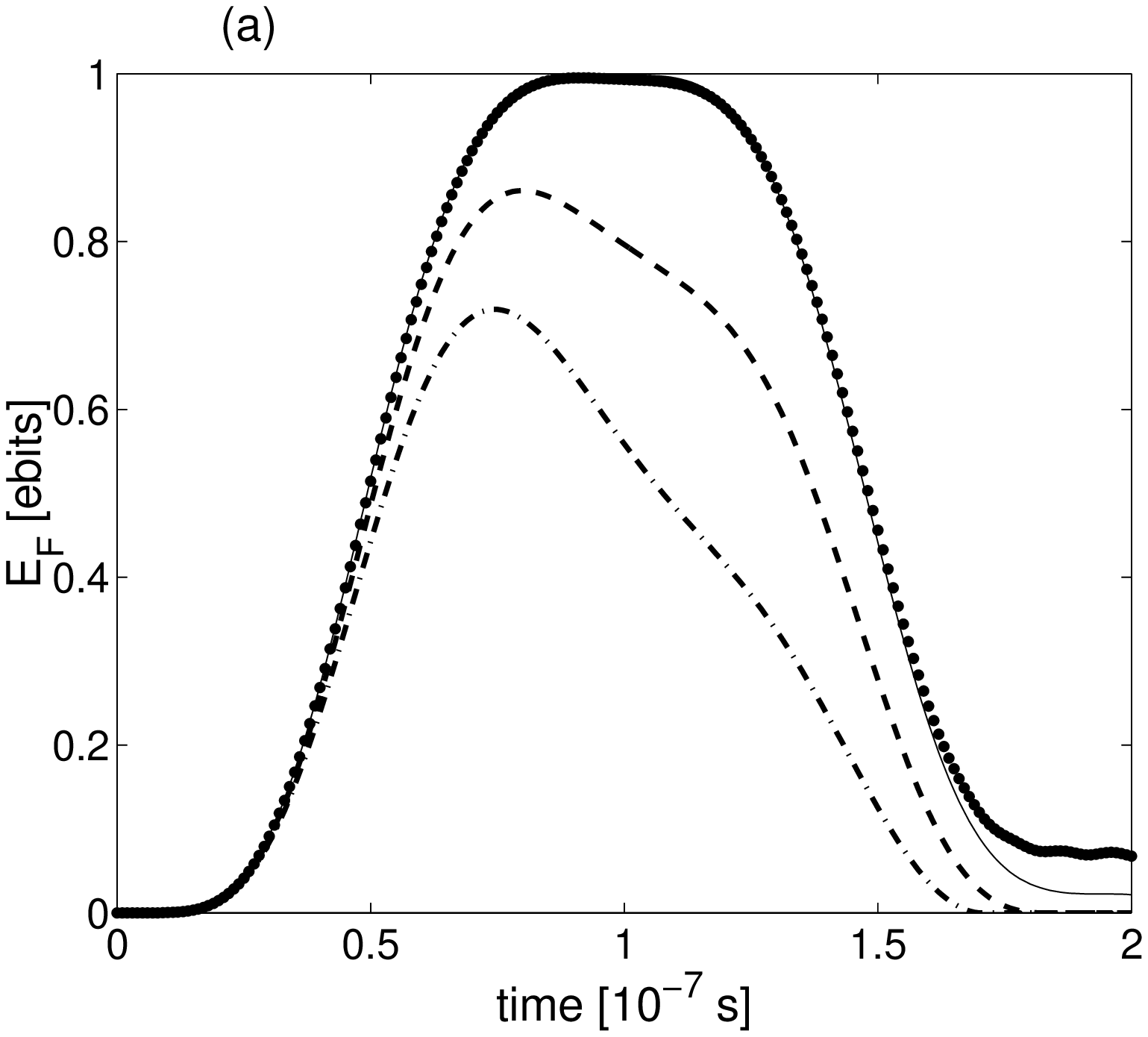}}
\epsfxsize=7cm\centerline{\epsfbox{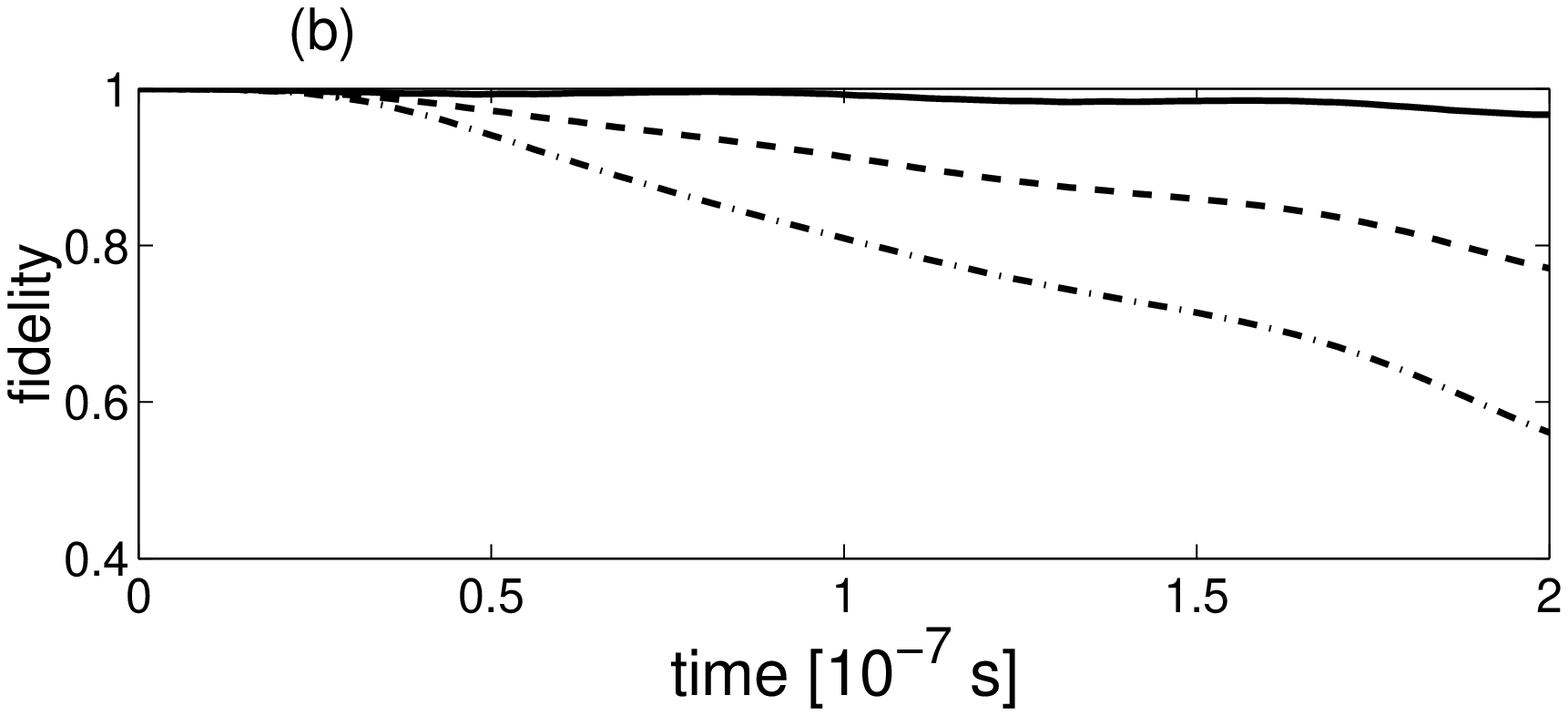}} \caption{Effect of
the standard damping, described by (\ref{N38}) and (\ref{N39}) for
quiet reservoirs, on (a) the entanglement of formation and (b)
fidelity $F(\hat{\rho},\hat{\rho}_{\rm cut})$, of the states
generated in the single-mode pumped coupler from the truncated
two-qubit mixed states for $\chi_a=\chi_b=10^8$rad/sec,
$\alpha=\chi_a/20$, $\epsilon=\alpha/2$, and the damping constants
$\gamma_a=\gamma_b$ equal to 0 (solid), $\chi_a/500$ (dashed), and
$\chi_a/200$ (dot-dashed curves). Large dots in figure (a)
correspond to the exact solution.}
\end{figure}
\begin{figure} 
\epsfxsize=7cm\centerline{\epsfbox{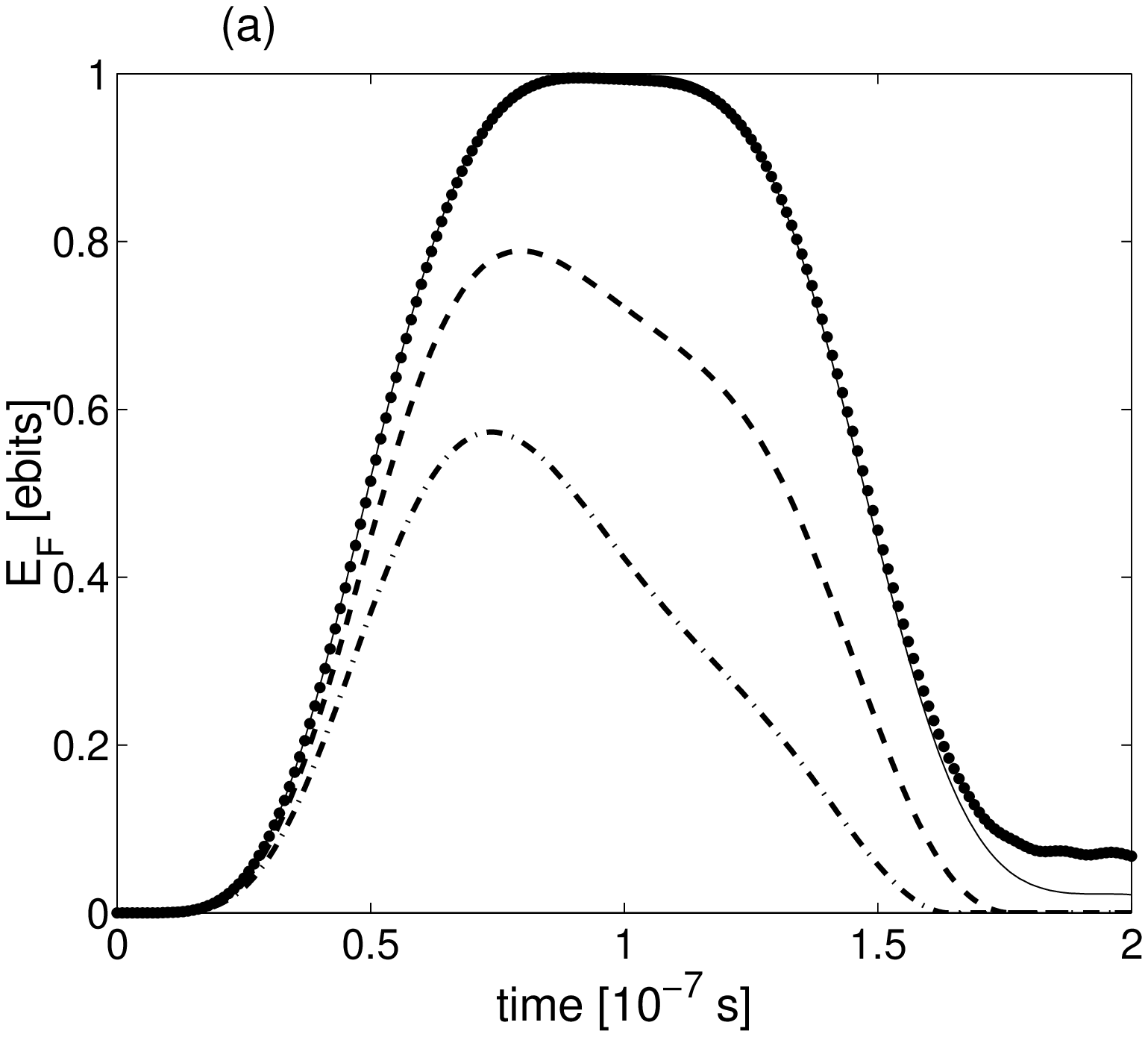}} %
\epsfxsize=7cm\centerline{\epsfbox{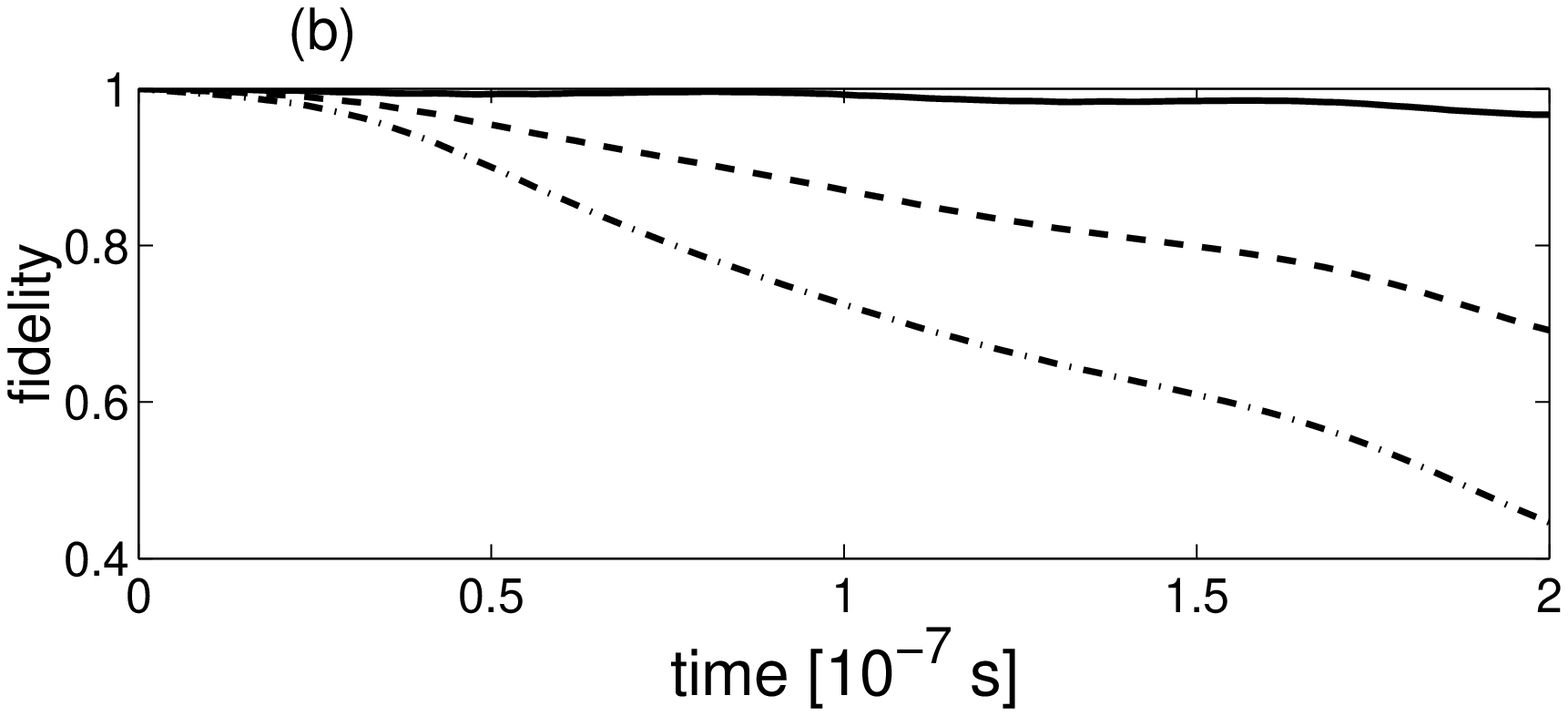}} %
\caption{Same as in figure 8 but for noisy reservoirs with mean
number $\bar n_a=\bar n_b=0.1$ of thermal photons.}
\end{figure}
\begin{figure} 
\epsfxsize=7cm\centerline{\epsfbox{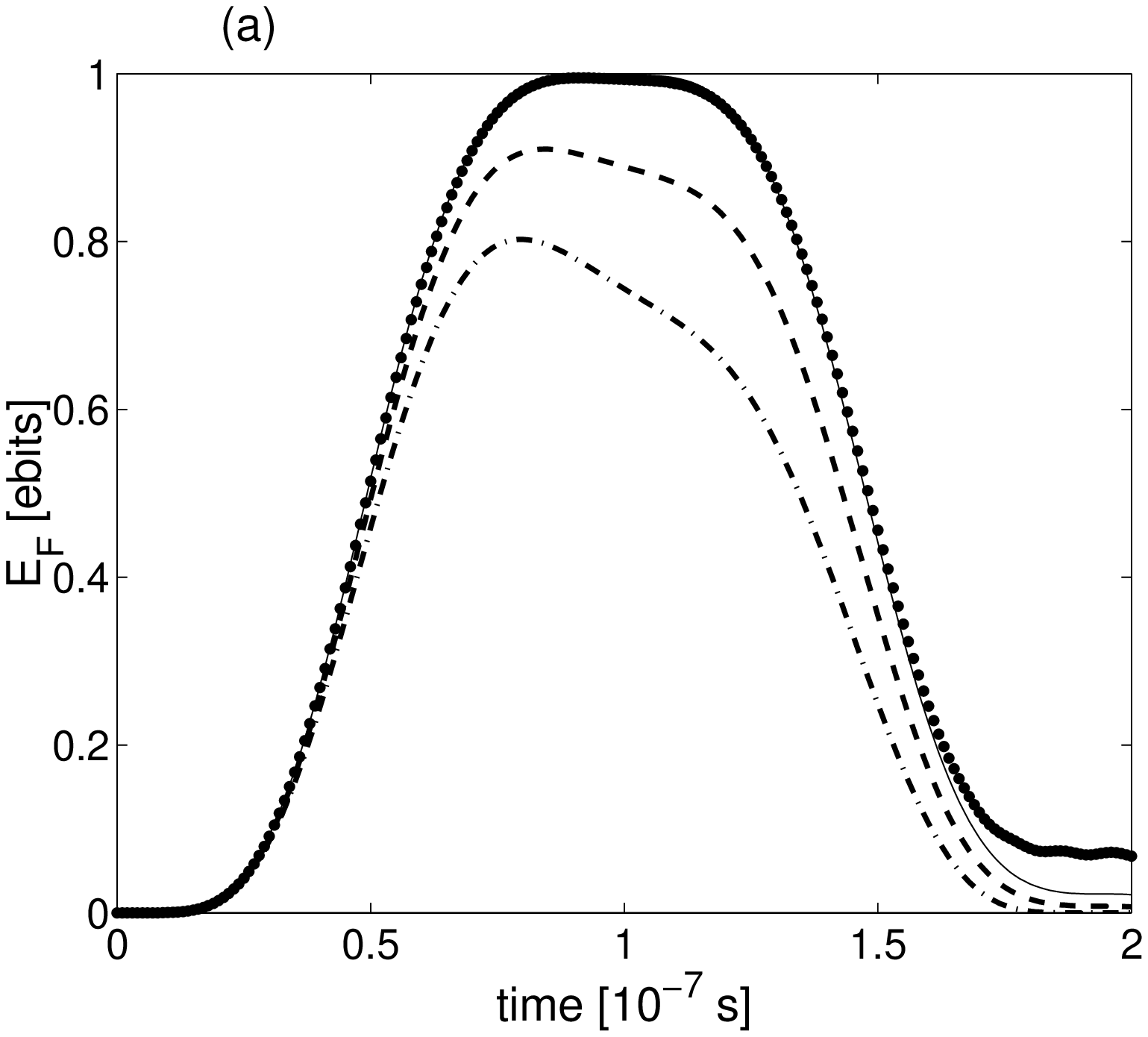}}
\epsfxsize=7cm\centerline{\epsfbox{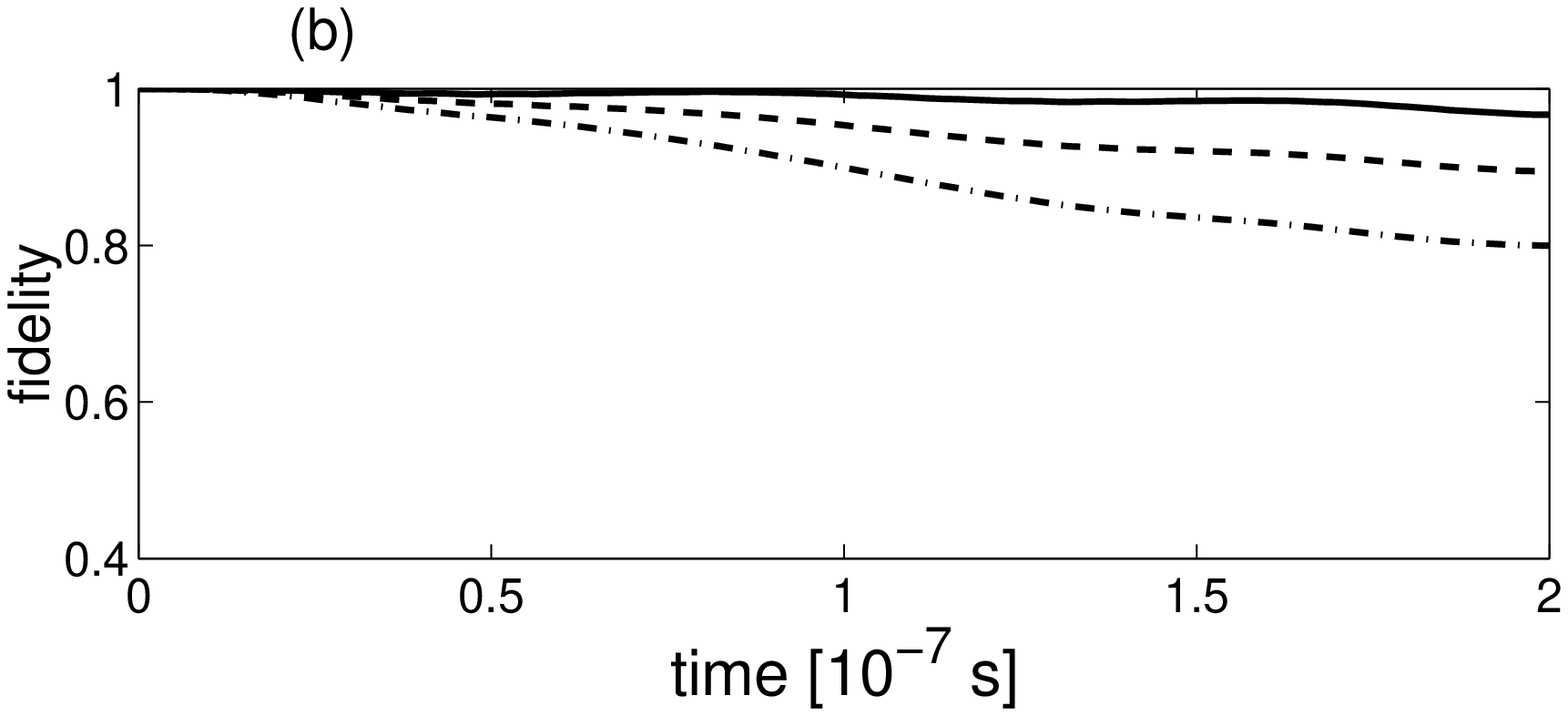}} \caption{Effect of
the phase damping described by (\ref{N38}) and (\ref{N41})
assuming quiet reservoirs for the same parameters as in figure~8.}
\end{figure}
\begin{figure} 
\epsfxsize=7cm\centerline{\epsfbox{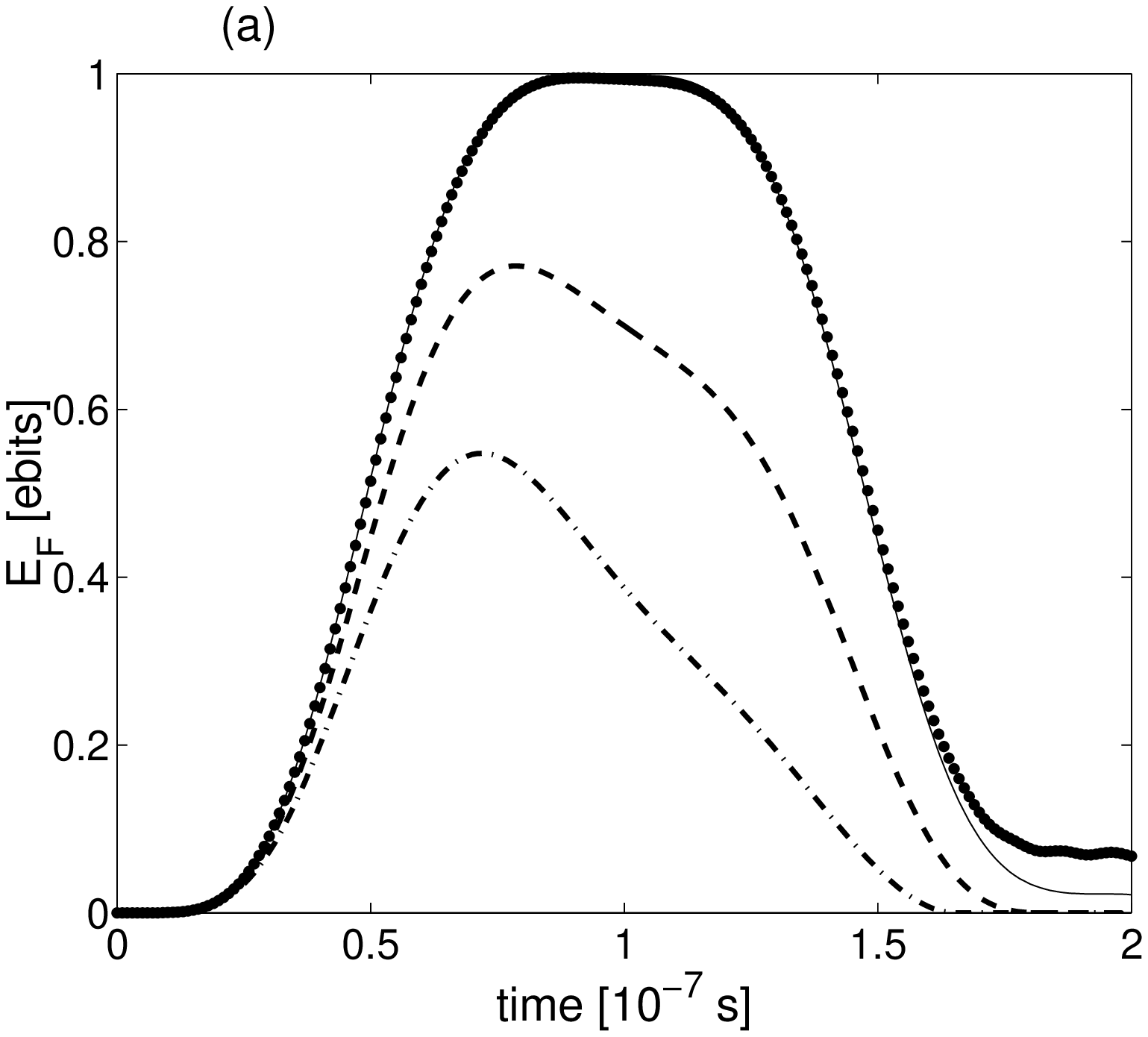}} %
\epsfxsize=7cm\centerline{\epsfbox{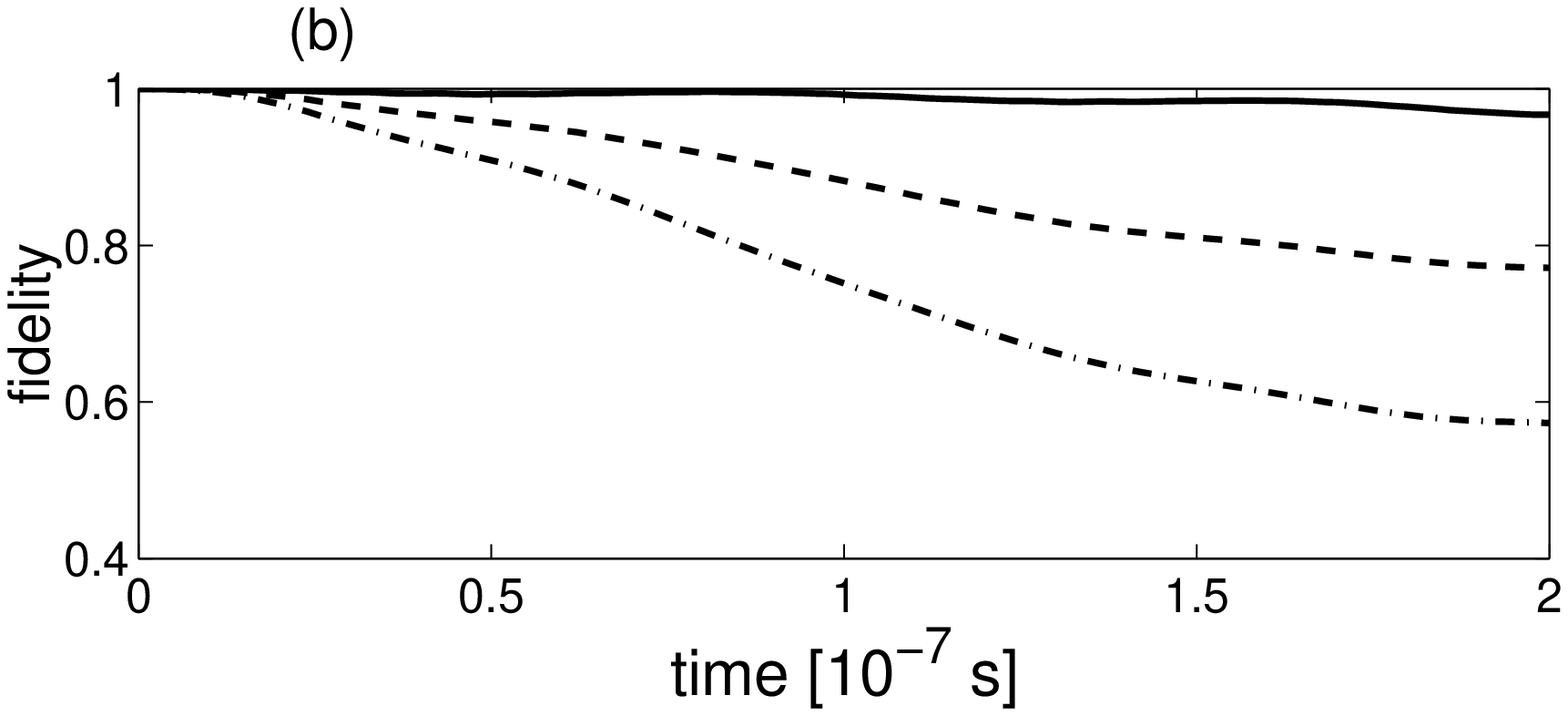}} %
\caption{Same as in figure 10 but for noisy reservoirs with $\bar
n_a=\bar n_b=1$, which is 10 times more than in figure 9.}
\end{figure}

The entropy of entanglement $E$, given by (\ref{N16}), is valid
for qudits of arbitrary dimension in a pure state, but it fails to
determine entanglement of a system in a mixed state. Thus, for a
two-qubit mixed state $\hat \rho$, we have to apply more general
measure, e.g., the Wootters measure of entanglement of formation
given by \cite{Woo98}:
\begin{equation}
E_F(\hat \rho)={\cal E}(C(\hat \rho)), \label{N42}
\end{equation}
where ${\cal E}$ is given by (\ref{N18}) with the argument being
the concurrence $C$ defined as
\begin{equation}
C(\hat \rho) =\max \{2\max_{i}\lambda _{i}-\sum_{i=1}^{4}\lambda
_{i},0\}, \label{N43}
\end{equation}
and $\lambda _{i}$ are the square roots of the eigenvalues of
$\hat{\rho}(\hat{\sigma}^{(a)}_{y}\otimes
\hat{\sigma}^{(b)}_{y})\hat{\rho} ^{\ast }(\hat{\sigma}^{(a)}_{y}
\otimes \hat{\sigma}^{(b)}_{y})$, while $\hat{\sigma}^{(k)} _{y}$
is the Pauli spin matrix of the $k$th qubit ($k=a,b$). It is
well-known that the entanglement of formation goes into the
entropy of entanglement for any two-qubit pure states. Examples of
evolution of the entanglement of formation and fidelity for the
dissipative systems are shown in figures 8--11 both for quiet and
noisy reservoirs. By analyzing the figures, the most important
observation is that the generated two-qubit entangled states are
very fragile to the leakage of photons from the cavities in the
analyzed couplers. Our system is more fragile to losses described
by the standard master equation, given by (\ref{N38}) and
(\ref{N39}), rather than the losses due to only phase damping as
described by the Gardiner-Zoller master equation, given by
(\ref{N38}) and (\ref{N41}). Inclusion of reservoir noise, at
least for the assumed low mean numbers $\bar n\equiv \bar n_a=\bar
n_b$ of thermal photons, does not cause a dramatic deterioration
of fidelity and entanglement in comparison to the losses caused by
coupling the system to the zero-temperature reservoirs. Note that
we have chosen $\bar n=1$ in the dephasing model shown in figure
11 and much smaller value of $\bar n$ in the standard dissipation
model in figure 9. The reason is that for the latter dissipation
model, the number of photons in the system can be increased by
absorbing thermal photons from the reservoirs. If this absorption
of thermal photons exceeded the loss of the system photons to the
reservoirs then the generated fields could not be approximated as
two-qubit states and the Wootters function $E_F$, given by
(\ref{N42}), would fail to be a good entanglement measure. For the
parameters chosen in figure 9, the generated states are well
approximated by two-qubit density matrices and thus its
entanglement of formation is well described by (\ref{N42}). By
contrast, the number of system photons is not affected by the
reservoirs in the dephasing model; thus (\ref{N42}) can be used
for any number $\bar n$ of thermal photons.

Finally, it should be noted that fragility of our system to
dissipation seems to be a serious drawback from an experimental
point of view. Nevertheless, a method which enables a significant
improvement of the entanglement robustness of the generated
states, has recently been suggested for a similar system
\cite{Leo05}.
\begin{figure} 
\hspace*{1cm}\epsfxsize=9cm\centerline{\epsfbox{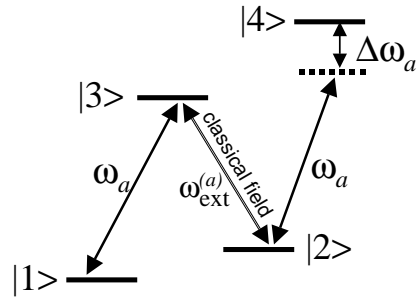}}
\caption{Level structure of atoms in the Schmidt-Imamo\v{g}lu
system \cite{Sch96,Ima97} exhibiting a resonantly enhanced Kerr
nonlinearity in mode $a$. An analogous figure can be drawn for
mode $b$.}
\end{figure}

\section{Discussion and conclusions}

One of the crucial conditions for the successful truncation and
generation of the Bell states in our scheme concerns the couplings
$|\alpha|$, $|\beta|$ and $|\epsilon|$ to be much smaller than the
Kerr nonlinearities $\chi_a$ and $\chi_b$. This implies that the
Kerr interaction should be strong at very low light intensities.
Thus, it is desirable to discuss an implementation, in which such
stringent conditions can experimentally be satisfied. A possible
realization can be based on the effect of the
electromagnetically-induced transparency (EIT) or atomic dark
resonances as proposed by Schmidt and Imamo\v{g}lu
\cite{Sch96,Ima97} (see also \cite{Gra98}) and observed
experimentally \cite{Hau99,Kan03}. The Schmidt-Imamo\v{g}lu EIT
scheme can be realized in a low density system of four-level
atoms, which level structure is shown in figure 12, exhibiting
giant resonantly-enhanced Kerr nonlinearity at very low
intensities. The atoms are placed in cavity $a$ (and analogously
in cavity $b$) tuned to frequency $\omega_a$ of the mode $a$
resonant with the transition $|1\rangle \leftrightarrow|3\rangle$
and detuned by $\Delta\omega_a$ of the transition
$|2\rangle\leftrightarrow|4\rangle$. The EIT effect is created by
a classical pumping field of frequency $\omega_{\rm ext}^{(a)}$
resonant with the transition $|2\rangle \leftrightarrow
|3\rangle$. By assuming $|g_{13}|^2 n_{\rm atom}/\Omega_{a}^{2}<1$
(see \cite{Gra98}), all the atomic levels can adiabatically be
eliminated, which results in the following formula for the Kerr
nonlinearity \cite{Ima97}:
\begin{equation}
2\chi_{a} \sim \frac{3\hbar \omega_{a}^2}{2\epsilon_0 V_a}{\rm Re}
(\chi_a^{(3)}) = \frac{3|g^{(a)}_{13}|^2|g^{(a)}_{24}|^2}{
\Omega^2_{a} \Delta\omega_a} n_{\rm atom},\label{N44}
\end{equation}
where $\chi_a^{(3)}$ is third-order nonlinear susceptibility,
$g_{ij}=\mu_{ij}\sqrt{\omega_i/(2\hbar \epsilon_0 V_a)}$ are the
coupling coefficients, $\Omega_{a}$ is the Rabi frequency of the
classical driving (and coupling) field, $\mu_{ij}$ is the electric
dipole matrix element between the states $|i\rangle$ and
$|j\rangle$, $n_{\rm atom}$ is the total number of atoms contained
in the cavity of volume $V_{a}$, and $\epsilon_0$ is the
permittivity of free space. By replacing subscript $a$ by $b$ in
(\ref{N44}), an analogous expression for $\chi_b$ can be obtained.
By putting the stringent limit on the required cavity parameters
\cite{Gra98}, Imamo\v{g}lu {\em et al.} estimated $\chi_{a}\sim
10^8$ rad/sec \cite{Ima97}. Note that some quantum information
applications of these giant Kerr nonlinearities have already been
studied \cite{Dua00,Vit00,Ott03,Baj04,Mir04}. More details about
application of the Schmidt-Imamo\v{g}lu scheme of
resonantly-enhanced Kerr nonlinearity in the analysis of
dissipation effects on entanglement generation are given by one of
us in \cite{Mir04a}.

It is worth stressing the differences between the present paper
and our former works. (i) In \cite{Leo04,Leo05}, only the
single-mode pumped systems were studied. Here, we analyze also
two-mode pumped systems, which is basically a different model.
(ii) The model described in \cite{Leo05} is crucially different
from the one used here as was based on the two-mode {\em
nonlinear} interaction term $\hat H_{\rm int}=\epsilon \hat
a^{2\dagger}\hat b^2+{\rm h.c.}$. In the present work, as well as
in \cite{Leo04}, we apply the two-mode {\em linear} interaction
Hamiltonian $\hat H_{\rm int}=\epsilon \hat a^{\dagger}\hat b+{\rm
h.c.}$, which is the same (by neglecting the pump terms $\hat
H^{(a)}_{\rm ext}$ and $\hat H^{(b)}_{\rm ext}$) as that used in
\cite{Che96,Hor89,Kor96,Fiu99,Ibr00,Ari00,San03,ElOrany05}. The
other novelties of the present work in comparison to
\cite{Leo04,Leo05} can be summarized as follows: (1) for
single-mode pumped system, we found a new generalized solution
(\ref{N12}), which, in a special case of equal couplings $\alpha$
and $\epsilon$, simplifies to our former solution obtained in
\cite{Leo04}. (2) Analytical approximate solutions were found here
for various initial Fock states $|n\rangle$. In
\cite{Leo04,Leo05}, solutions were given for initial vacuum states
only. (3) Here, we describe a possible realization of the model
based on the effect of the electromagnetically-induced
transparency. (4) In present numerical analysis based on the EIT
scheme, in comparison to \cite{Leo04,Leo05}, more realistic
parameters were chosen for the coupling constants, Kerr
nonlinearities, and damping constants. (5) The deterioration of
the fidelity of the generated Bell states due to the standard
dissipation and phase damping was analyzed here within two types
of master equations both for quiet and noisy reservoirs. By
contrast, analysis of losses in \cite{Leo04} was limited to the
standard master equation and for the quiet reservoir only. No
effect of dissipation was studied in \cite{Leo05}. (6) In the
present manuscript, entanglement of formation was calculated
analytically for dissipation-free systems and calculated
numerically for dissipative systems. No analytical formulae were
given in \cite{Leo04}, while the entanglement of formation was not
at all studied in \cite{Leo05}. (7) The quality of truncation was
described here, but it was not studied in \cite{Leo04,Leo05}.
Discrepancy between the really generated states and the exact
truncated states was measured here by the fidelity.

In conclusion, we have described a realization of the generalized
two-mode optical-state truncation of two coherent modes via a
nonlinear process. Our system is a two-mode generalization of the
single-mode nonlinear quantum scissors device described in
\cite{Leo97}. We have described an implementation of the Kerr
nonlinear couplers, where resonantly enhanced nonlinearities can
be achieved in the Schmidt-Imamo\v{g}lu EIT scheme. We have
compared Kerr nonlinear couplers linearly excited in one or two
modes by external classical fields. We have shown under assumption
of the coupling strengths to be much smaller than the nonlinearity
parameters $\chi_a$ and $\chi_b$ that the optical states generated
by the couplers are the two-qubit truncated states spanned by
vacuum and single-photon states. Although our approximate
solutions of the Schr\"odinger equation are independent of
$\chi$-constants, the Kerr nonlinearity plays a crucial role in
the physics as we have derived from the complete $\chi$-dependent
Hamiltonian. In fact, the Kerr interaction is the mechanism in our
model, which enables truncation of the generated state at some
energy level. By contrast, the system without the Kerr
nonlinearities and pumped by an external field would gain more and
more energy. To confirm our predictions, we have compared `exact'
(accurate up to double-precision) direct numerical solutions of
the Schr\"odinger equation and compared with our approximate
analytical solutions. The discrepancies between the exact and
approximate solutions are relatively small as shown by fidelities
in figures 2--4. We have demonstrated that our system initially in
vacuum state or single-photon Fock states $|mn\rangle$ ($m,n=0,1$)
evolves into Bell states. We have discussed the fragility of the
entanglement of formation of the generated states due to the
standard dissipation and dephasing in the two distinct master
equation approaches.

\begin{acknowledgments}
We thank Professor Ryszard Tana\'s, Professor Ryszard Horodecki
and Professor Robert Alicki for discussions. This work was
supported by the Polish State Committee for Scientific Research
under grant No. 1 P03B 064 28.
\end{acknowledgments}

\end{document}